\documentclass[aps, prl, amsmath, superscriptaddress,preprintnumbers,twocolumn]{revtex4-2}
\usepackage{graphicx}
\usepackage{dcolumn}
\usepackage{bm}
\usepackage{amsmath}
\usepackage{subfigure}
\usepackage{amsfonts}
\usepackage[svgnames]{xcolor}
\usepackage{appendix}
\usepackage{multirow}
\usepackage{booktabs}
\usepackage{tabularx}
\usepackage[colorlinks,
            linkcolor=blue,
			filecolor=blue,
			urlcolor= blue,
			citecolor=blue,
            ]{hyperref}
\usepackage{amsthm}
\usepackage{setspace}
\usepackage[T1]{fontenc}
\usepackage{extarrows}
\usepackage{lineno}

\usepackage[percent]{overpic}
\usepackage{dsfont}
\usepackage{floatrow}
\usepackage{physics}
\usepackage{CJKutf8}

\newcommand{\hket}[1]{\ket{#1}\rangle}  
\newcommand{\hbra}[1]{\langle\bra{#1}}  

\newcommand{\kett}[1]{\left.\ket{#1}\right\rangle}
\newcommand{\braa}[1]{\left\langle\bra{#1}\right.}
\newcommand{\brakett}[2]{\left\langle\braket{#1}{#2}\right\rangle}

\renewcommand{\I}{\mathbb{I}}
\newcommand{\phii}[0]{\hat{\Phi}}

\newcommand{\M}[0]{{\mathcal{M}}}

\newcommand{\prlsection}[1]{{\em {#1}.---~}}

\begin{document}
\title{Quantum Zeno and Anti-Zeno Responses: Universal Spectral Criterion for Measurement-Induced Decay}

\author{Yu-Xiang Chen  }
\thanks{These authors contributed equally to this work.}
\affiliation{State Key Laboratory of Semiconductor Physics and Chip Technologies, Institute of Semiconductors, Chinese Academy of Sciences, Beijing 100083, China}
\affiliation{Center of Materials Science and Opto-Electronic Technology, University of Chinese Academy of Sciences, Beijing 100049, China}

\author{Yuan-De Jin }
\thanks{These authors contributed equally to this work.}
\affiliation{State Key Laboratory of Semiconductor Physics and Chip Technologies, Institute of Semiconductors, Chinese Academy of Sciences, Beijing 100083, China}
\affiliation{Center of Materials Science and Opto-Electronic Technology, University of Chinese Academy of Sciences, Beijing 100049, China}

\author{Wen-Long Ma}
\email{wenlongma@semi.ac.cn}
\affiliation{State Key Laboratory of Semiconductor Physics and Chip Technologies, Institute of Semiconductors, Chinese Academy of Sciences, Beijing 100083, China}
\affiliation{Center of Materials Science and Opto-Electronic Technology, University of Chinese Academy of Sciences, Beijing 100049, China}

\date{\today }

\begin{abstract}
We develop a general framework for characterizing the response of an evolving quantum system to repetitive quantum measurements. Modeling each evolution-measurement cycle as a quantum channel induced by an effective Liouvillian generator, we find that the Liouvillian spectral gap determines the measurement-induced decay rate. We analyze how the spectral gap responds to the measurement frequency, and define a quantum Zeno response as a decrease in the gap with increasing measurement frequency, and an anti-Zeno response as the opposite. We illustrate this criterion for both discrete- and continuous-time quantum measurements. In an exactly solvable discrete‑time qubit model, the exceptional‑point spectral coalescence or spectral crossings mark the transition between Zeno and anti-Zeno responses, which can be experimentally distinguished from the long-time
decay envelope of the survival probability. In a continuous‑time superconducting‑qubit–defect model, the same transition manifests as  smooth extrema of the spectral gap. Our results establish a universal spectral criterion for measurement-induced decay, offering a practical route to identify and manipulate these effects in generic quantum systems.
\end{abstract}
\maketitle

\prlsection{Introduction}
Quantum Zeno and anti-Zeno effects describe measurement-induced modification of quantum dynamics \cite{Itano1990,Schulman1998,Facchi2001,Facchi2009,Kofman2000,Kofman2001,Zheng2008,Facchi2008}. In unstable or open quantum systems, they are conventionally identified as suppression or enhancement of decay relative to the unmeasured evolution; in closed quantum systems, repeated noncommuting measurements similarly slow or accelerate coherent mixing. These phenomena, underpinning applications in entanglement or state protection \cite{Dhar2006,Maniscalco2008,Bernu2008,silva2012}, quantum control \cite{Signoles2014,Hacohen-Gourgy2018,Lewalle2024}, error correction \cite{Erez2004,silva2012,Ofek2016} and sensing \cite{Macieszczak2015,Virzi2022,Ronchi2024,Muller2016,Long2022}, have been observed in diverse platforms, such as trapped ions \cite{Itano1990,Balzer2002,Zhang2018b}, cold atoms \cite{Fischer2001,Streed2006,Schafer2014,Patil2015,Peise2015}, superconducting circuits \cite{Kakuyanagi2015,Slichter2016,Harrington2017,Thorbeck2024}, solid-state spin systems \cite{Wolters2013,Kalb2016,Dasari2022}, and acoustic waveguides \cite{Zhang2025}.

Placing these effects on a common footing is challenging. A conventional, zeroth-order criterion compares the measurement-modified decay rate with the natural one \cite{Facchi2001,Cao2010,BhaktavatsalaRao2011,Zhang2018}, but this comparison becomes ambiguous when a closed system or a more complex system has no unique intrinsic decay rate. 
Recently, a local frequency-response criterion has classified the dynamics through the derivative of an effective decay rate with respect to measurement frequency \cite{Greenfield2025}. 
Such a criterion has been applied to quantum dephasing dynamics \cite{Chaudhry2014} and relaxation processes in spin baths and structured reservoirs \cite{Segal2007,Chaudhry2016,Wu2017}. It has also revealed connections between Zeno dynamics and parity-time-symmetry breaking \cite{Li2023} or strong-coupling dissipation \cite{Khan2022}. On the other hand, under continuous partial measurements, the decay rate can also be tuned through noise or measurement backaction \cite{Koshino2005,Layden2015,Snizhko2020,Kumar2020,Chen2024}.
These findings establish important criteria in their respective settings; however, a comprehensive response theory for quantum Zeno and anti-Zeno effects is still lacking, raising two key open problems: Can we build a unified framework to quantitatively characterize the response of a generic (closed or open) quantum system to repeated measurements of arbitrary strength? Can such a characterization be connected to the spectral properties of the quantum channel defining a single evolution-measurement cycle?

\begin{figure}
    \centering
    \includegraphics[width=\linewidth]{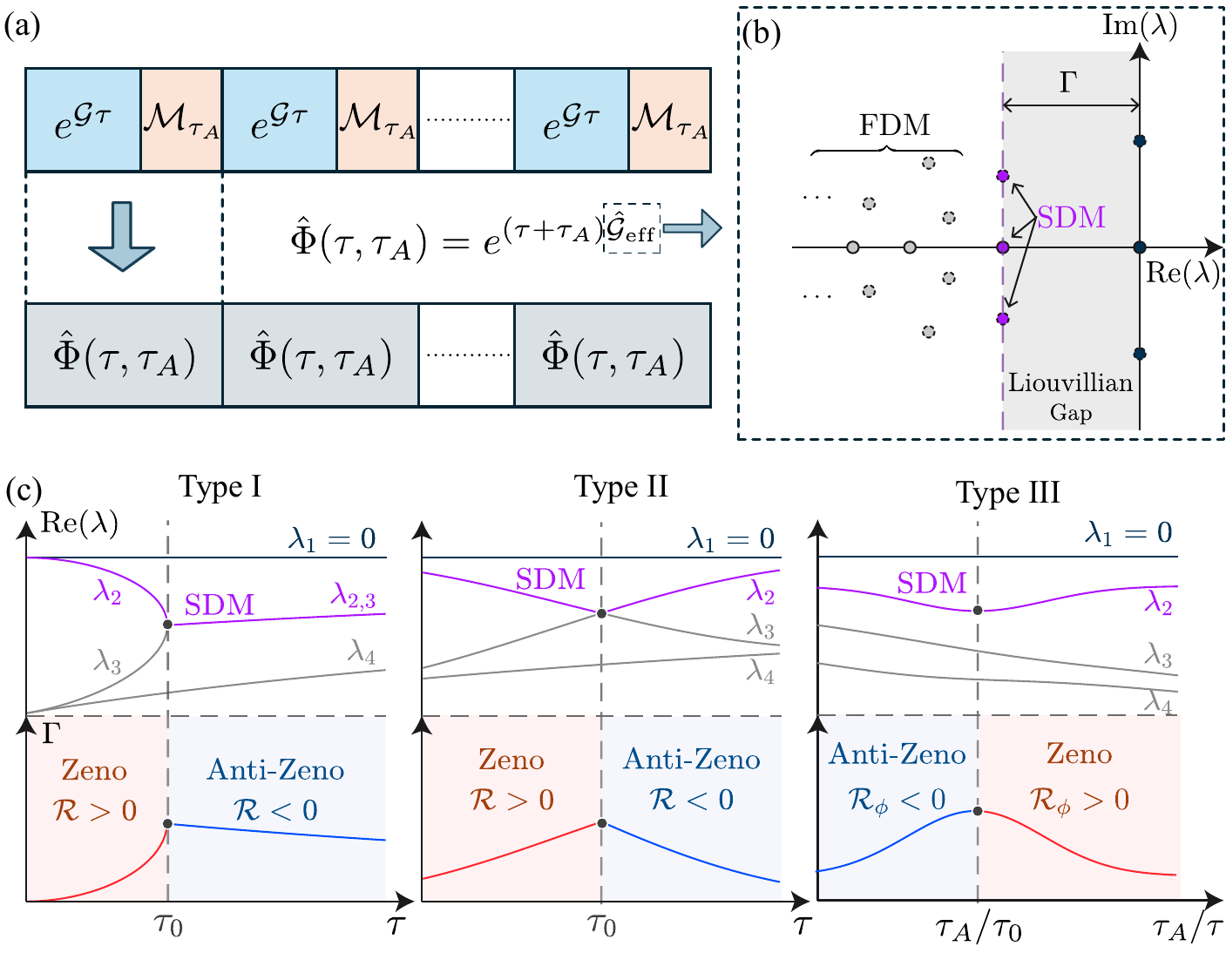}
    \caption{Schematic illustration of quantum Zeno and anti-Zeno responses. (a) Each cycle contains a free evolution generated by $\mathcal{G}$ for time $\tau$ followed by a generalized measurement $\M_{\tau_A}$ for time $\tau_A$, modeled by a concatenated channel induced by an effective generator $\mathcal{G}_{\rm eff}$. 
    (b) The eigenmodes of $\hat{\mathcal{G}}_{\rm eff}$ are divided into fast decay modes (FDMs) and SDMs. The decay behavior of the system is governed by the Liouvillian gap $\Gamma$, corresponding to the absolute value of the real part of SDM. (c) The Zeno-to-anti-Zeno response transitions (with a fixed $\tau_A$) can occur for three types of spectral transitions: a spectral coalescence at an EP (Type-I), a spectral crossing of competing decay modes (Type-II), or a smooth extremum of $\Gamma$ (Type-III).
    }
    \label{fig:placeholder} 
\end{figure}

In this paper, we provide a general framework to characterize the response of generic quantum dynamics to repetitive generalized quantum measurements. By formulating each cycle of evolution and measurement as a quantum channel [Fig.~\hyperref[fig:placeholder]{\ref*{fig:placeholder}(a)}], we derive an effective Liouvillian generator with its spectral gap \cite{Macieszczak2016,Minganti2018} determining the decay rate of the slowest decay mode (SDM) [Fig.~\hyperref[fig:placeholder]{\ref*{fig:placeholder}(b)}]. We define a quantum Zeno response when the derivative of the spectral gap with respect to the measurement frequency is negative, and an anti-Zeno response conversely.  
Within this spectral-response formulation, a change between Zeno and anti-Zeno responses can arise through several distinct spectral mechanisms [Fig.~\hyperref[fig:placeholder]{\ref*{fig:placeholder}(c)}]: a spectral coalescence at an exceptional point (EP), a spectral crossing between competing decay modes, or a smooth extremum of the spectral gap. We first make this classification explicit in an analytically solvable closed-system model, where a single qubit undergoes unitary precession interrupted at discrete times by generalized measurements implemented through Ramsey measurements on an ancilla. This exactly solvable model shows how varying the measurement interval and strength drives Zeno-to-anti-Zeno transitions through spectral coalescence or spectral crossings. We then move to a continuous-time measurement setting motivated by recent superconducting-circuit experiments \cite{Thorbeck2024}, where a superconducting qubit is coupled to a lossy two-level-system defect and the measurement is modeled as dephasing. Applying the same spectral criterion, we recover the measurement-dependent decay rate and show that the response boundary can appear as a smooth extremum of the spectral gap. Our findings therefore provide new perspectives on unifying quantum Zeno and anti-Zeno effects in closed or open quantum systems under discrete- and continuous-time quantum measurements.

\prlsection{Quantum Zeno and anti-Zeno responses}
We first construct a general model for defining the quantum Zeno and anti-Zeno responses. We consider a $d$-dimensional quantum system that undergoes quantum evolution governed by a generator $\mathcal{G}$, while being periodically measured at intervals of $\tau$. Then each evolution-measurement cycle can be described by a concatenated quantum channel
\begin{equation}
\Phi(\tau,\tau_{A})=\mathcal{M}_{\tau_{A}} e^{\mathcal{G}\tau},
\end{equation}
 where $\mathcal{G}=-i[H,\cdot]+\sum_{k=1}^r \gamma_k\mathcal{D}[L_k]$ is the dynamical generator, with $\mathcal{D}[L_k]=L_k(\cdot)L_k^\dagger -\frac{1}{2}\{L_k^\dagger L_k, (\cdot)\}$. Here $\gamma_k$ and $L_k$ denote the relaxation rate and the corresponding jump operator, respectively.  The generator $\mathcal{G}$ can describe either unitary dynamics of a closed system or Markovian dynamics of an open quantum system. The measurement channel $\mathcal{M}_{\tau_{A}}(\cdot)=\sum_\alpha M_{\tau_{A},\alpha}(\cdot) M_{\tau_{A},\alpha}^\dagger$ represents a measurement taking time $\tau_A$, where the Kraus operators $\{M_{\tau_{A},\alpha}\}$ satisfy $\sum_\alpha M_{\tau_{A},\alpha}^\dagger M_{\tau_{A},\alpha}=\mathbb{I}$ with $\mathbb{I}$ being the identity operator.


The system dynamics under such sequential measurements can be uncovered by examining the spectral characteristics of $\Phi(\tau,\tau_{A})$. We represent it as a matrix $\hat{\Phi}$ in the vectorized operator space, where a superoperator $X(\cdot)Y$ is represented by $X\otimes Y^T$ under the vectorization $R=\sum_{mn}r_{mn}\ket{m}\bra{n}\to\hket{R}=\sum_{mn}r_{mn}\ket{m}\otimes\ket{n}$, with $(\cdot)^T$ denoting matrix transposition.
If the channel is diagonalizable, we can spectrally decompose $\hat{\Phi}$ as [see Sec. S1 of Supplementary Material (SM) \footnote{See Supplementary Material for more analysis and calculations about both discrete- and continuous-time measurement models} for details]
\begin{equation}
\hat{\Phi} (\tau,\tau_{A})=e^{\hat{\mathcal{G}}_{\rm eff}(\tau+\tau_A)}=\sum_{i=1}^{d^2}e^{\lambda_i(\tau+\tau_A)}\hket{R_i}\hbra{L_i},
\end{equation}
where $\{\lambda_i\}$ denotes the spectrum of an effective generator $\hat{\mathcal{G}}_{\rm eff}$, and $\{\hket{R_i},\hket{L_i}\}$ is a complete biorthonormal system satisfying $\brakett{L_i}{R_j}=\delta_{ij}$. For weak measurements, we can approximate the measurement channel by $\hat{\mathcal{M}}_{\eta}\approx e^{\hat{\mathcal{G}}_{m}\tau_{A}}$. Thus, for short $\tau$ and $\tau_A$ such that $\| [\hat{\mathcal{G}},\hat{\mathcal{G}}_{m}] \|\tau\tau_{A} \ll 1$, we have $\Phi(\tau,\tau_{A})=e^{\hat{\mathcal{G}}_{m}\tau_{A}}e^{\hat{\mathcal{G}}\tau}\approx e^{\hat{\mathcal{G}}_{\mathrm{eff}}(\tau+\tau_{A})}$ with $\hat{\mathcal{G}}_{\mathrm{eff}} = (\tau\hat{\mathcal{G}} + \tau_{A}\hat{\mathcal{G}}_{m}) / (\tau+\tau_{A})$ \cite{Trotter1959,Lerner2018,Han2021}. This corresponds to the continuous-time measurement models \cite{Chaudhry2014,Streed2006,Jacobs2006}, where the measurement is modeled as additional dissipative dynamics.

We order the eigenvalues of $\hat{\mathcal{G}}_{\rm eff}$ by decreasing real parts with ${\rm Re}(\lambda_{i+1})\leq{\rm Re}(\lambda_{i})\leq 0$, in which the modes with $\Re(\lambda_i)=0$ span the invariant subspace. As exemplified below, the invariant subspace is often independent of $\tau$ and $\tau_A$. Let $s$ be the dimension of the invariant subspace, then the leading real eigenvalue $\lambda_{s+1}$ (or a complex conjugate pair with $\lambda_{s+1}=\lambda_{s+2}^{*}$) outside this subspace corresponds to the SDM, and the corresponding decay rate is determined by the Liouvillian spectral gap $\Gamma=|{\rm Re}(\lambda_{\rm SDM})|=|{\rm Re}(\lambda_{s+1})|$. 

\begin{figure*}[htbp]
    \centering
    \includegraphics[width=\textwidth]{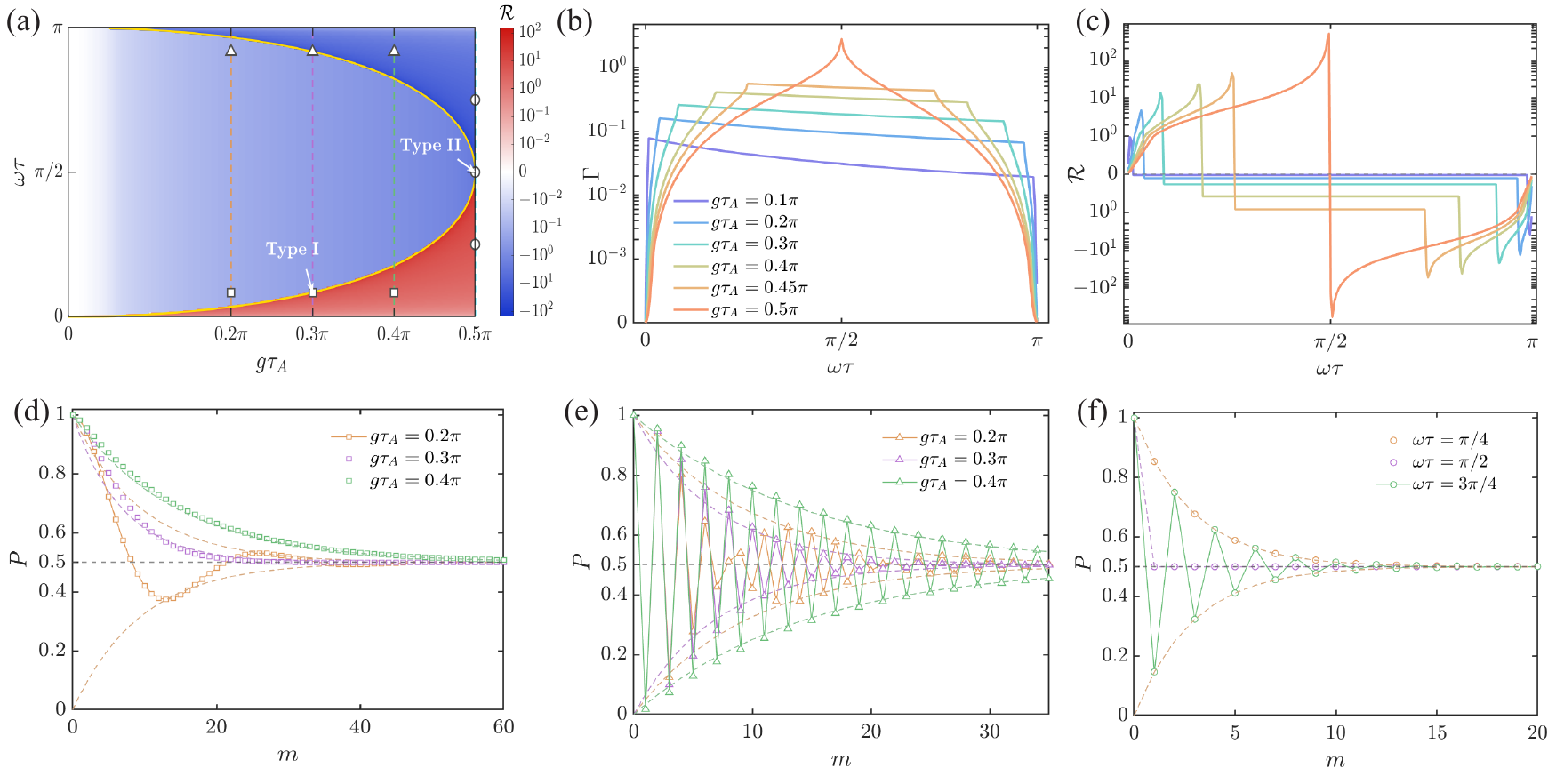}
    \caption{Quantum Zeno and anti-Zeno responses of a unitarily evolving qubit to repetitive discrete-time measurements with tunable strength. (a) A phase diagram of the response $\mathcal{R}$ in the ($g\tau_A, \omega\tau$) parameter plane. The lower half of the EP line (yellow curve) demarcates the boundary between the anti-Zeno response ($\mathcal{R} < 0$) and Zeno response ($\mathcal{R} > 0$). Vertical dashed lines mark different measurement strengths $g\tau_A$ corresponding to the cross-sections in (b) and (c). (b) The decay rate $\Gamma$ of the SDM as a function of the evolution time $\omega\tau$ for various $g\tau_A$. Maxima of $\Gamma$ coincide with EPs, which herald a transition in the underlying decay dynamics. (c) The response $\mathcal{R}$ of the spectral gap as a function of $\omega\tau$ for the same set of $g\tau_A$. The Zeno-to-anti-Zeno response transitions are governed by either spectral coalescence at EPs or spectral crossings (e.g., at $\omega\tau = \pi/2$ for $g\tau_A = 0.5\pi$). (d-f) Survival probabilities (represented by markers) of the qubit with the parameters denoted by the same markers in (a), where the dashed lines denote the decays predicted by spectral gaps. The other parameters are $\omega=\pi$ and $g=1$. }
    \label{fig:closesystem}
\end{figure*}

The decay behaviors of the system can be mostly captured by the SDM, as the system state after $m$ cycles can be represented as 
\begin{equation}
    \phii^m(\tau,\tau_{A})\kett{\rho}\approx\sum_{i=1}^s c_i\kett{R_i} +c_{s+1}e^{-\Gamma m(\tau+\tau_A)}\kett{R_{s+1}},
\end{equation}
where $c_i=\braa{L_i}\rho\rangle\rangle$ is a superposition coefficient. 
We define a response of the spectral gap to the measurement frequency $f=1/(\tau+\tau_A)$ with a fixed $\tau_A$ as \cite{Greenfield2025}
\begin{equation}
    \mathcal{R}(\tau,\tau_{A})=-\pdv{\Gamma}{f}=(\tau+\tau_A)^2\pdv{\Gamma}{\tau}.
\end{equation}
A positive $\mathcal R$ means that increasing the measurement frequency decreases the decay rate and hence suppresses the relaxation to the fixed point, giving a Zeno response. Conversely, a negative $\mathcal R$ indicates an anti-Zeno response. For continuous-time measurement models, the distinct timescales $\tau$ and $\tau_{A}$ are ill-defined, but we can still define the response as $\mathcal{R}_{\phi}=-\pdv{\Gamma}{\gamma_\phi}$ with $\gamma_{\phi}$ being proportional to the ratio $\tau_{A}/\tau$ (see the End Matter for details).

\prlsection{Discrete-time measurement models}
We first apply the criteria to an exactly solvable model, where a qubit continuously evolves interleaved periodically by a generalized measurement \cite{Pfender2019}. 
We consider a unitary evolution with a Hamiltonian $H=\omega\sigma_z/2$ so that $\hat{\mathcal{G}}=-i(H\otimes \mathbb{I}-\mathbb{I}\otimes H^{*})$. The generalized measurement can be implemented by a Ramsey interferometry measurement (RIM) on an ancilla qubit.
In each RIM, the ancilla qubit is first initialized to $\ket{0}_q$, rotated by a $\pi/2$-pulse $R_{\phi_1}=e^{-i(\cos\phi_1\sigma_q^x+\sin\phi_1\sigma_q^y)\pi/4}$ ($\sigma_q^i$ denotes the Pauli-$i$ operator of the ancilla with $i=x,y,z$), coupled to the measured qubit through a Hamiltonian $\sigma_q^z\otimes A$ with $A=g\sigma_x/2$ for a time $\tau_A$, and finally measured in the $z$-basis after a second $\pi/2$-pulse $R_{\phi_2}$. The measurement outcome $\alpha\in\{0,1\}$ corresponds to the Kraus operator $M_\alpha=[U_{+}-(-1)^\alpha e^{i\Delta\phi}U_{-}]/2$, and the measurement channel is $\hat{\mathcal{M}}_{\tau_A}=(U_+\otimes U_+^*+U_-\otimes U_-^*)/2$ with $U_{\pm}=e^{\mp ig\sigma_x\tau_A/2}$ and $\Delta \phi=\phi_1-\phi_2$. Upon choosing $\Delta \phi=\pi/2$, the Kraus operator can be written as $M_\alpha=[c_+\mathbb{I}+(-1)^\alpha c_-\sigma_x]/\sqrt{2}$ with $c_\pm=\sqrt{[1\pm\cos(g\tau_{A})]/2}$, and thus 
the measurement can be continuously tuned from weak to projective by varying $\tau_A$ from 0 to $\pi/(2g)$. Specifically, in the weak-measurement regime with $g\tau_{A}\ll 1$, the measurement channel becomes ${\mathcal{M}}_{\tau_{A}}\approx e^{{\mathcal{G}}_{m}\tau_{A}}$, where ${\mathcal{G}}_{m}=\gamma_{\phi}^{\rm eff}{\mathcal{D}}[\sigma_{x}]$ with $\gamma_{\phi}^{\rm eff}=g^{2}\tau_{A}/4$.


The eigenvalues of $\hat{\Phi}(\tau,\tau_{A})=e^{\hat{\mathcal{G}}_{\rm eff}(\tau+\tau_A)}$
can be exactly solved as $e^{\lambda_1(\tau+\tau_{A})}=1$, $e^{\lambda_{2,3}(\tau+\tau_{A})}=\cos^2\left(\frac{g\tau_A}{2}\right)\left[\cos(\omega\tau)\pm\sqrt{\tan^4\left(\frac{g\tau_A}{2}\right)-\sin^2(\omega\tau)}\right]$, and  $e^{\lambda_4(\tau+\tau_{A})}=\cos(g\tau_A)$. We can prove that independent of $\tau$ and $\tau_A$, the only fixed point of this channel is the maximally mixed qubit state (see Sec. S2A of SM). Below we show that the boundaries between the Zeno and anti-Zeno responses are attributed to either spectral coalescence at the EPs \cite{Heiss2012} or spectral crossings.

An exceptional line is determined by the condition $\tan^4(g\tau_A/2)=\sin^2(\omega\tau)$ [see the yellow line in Fig.~\hyperref[fig:closesystem]{\ref*{fig:closesystem}(a)}]. We consider a generalized measurement with $g\tau_A<\pi/2$, and gradually increase $\tau$ to examine the SDMs. With $\tau$ below the lower segment of the exceptional line, the SDM corresponds to a real eigenmode with $\lambda_2$, so that $\Gamma=|\rm Re(\lambda_{2})|$ and $\mathcal{R}>0$, corresponding to a Zeno response. As $\tau$ increases above lower segment, the SDMs become a pair of complex conjugate modes with 
$\Gamma=-\mathrm{ln}(\sqrt{|\cos(g\tau_A)|})/(\tau+\tau_A)$ and $\mathcal{R}=\mathrm{ln}(\sqrt{|\cos(g\tau_A)|})<0$, 
corresponding to an anti-Zeno response [see the plateau region in Fig.~\hyperref[fig:closesystem]{\ref*{fig:closesystem}(c)}]. Increasing $\tau$ further above the upper segment of the exceptional line results in a singularity of the anti-Zeno response. So the lower segment of the exceptional line not only signals the crossover between the Zeno and anti-Zeno responses, but also pinpoints the condition for maximal mixing rates [Fig.~\hyperref[fig:closesystem]{\ref*{fig:closesystem}(b)}]. 

In addition to spectral coalescence at EPs, spectral crossings also mark the boundaries between the Zeno and anti-Zeno responses. For projective measurements with $g\tau_A=\pi/2$, the paired eigenvalues $\lambda_2$ and $\lambda_3$ have identical moduli, resulting in a discontinuity in $\mathcal{R}$ and thereby marking the boundary between the Zeno and anti-Zeno responses. This finding provides a spectral explanation for the results in Ref. \cite{Li2023}. In Secs. S2B and S2C of the SM, we also perform extensive analysis and calculations to reveal the Zeno and anti-Zeno responses in open qubit and multi-qubit models. 

\prlsection{Observable signatures of quantum Zeno and anti-Zeno responses}
The quantum Zeno and anti-Zeno responses can be distinguished from the long-time decay envelope of the survival probability as the measurement frequency is varied. We define the survival probability for initial pure state $\rho_0=\ket{\psi_0}\bra{\psi_0}$ as $P(m)=\Tr[\rho_0\rho_m]$, with $\rho_m=\Phi^m(\rho_0)$. In the vectorized operator space, this quantity can be written as
\begin{equation}
  P(m)=\braa{\rho_0}\phii^m\kett{\rho_0}=\sum_{i=1}^{d^2}d_ie^{m\lambda_i(\tau+\tau_A)},
  \label{eq:pm_prob}
\end{equation}
with $d_i=\langle\langle{\rho_0}|R_i\rangle\rangle \langle\langle{L_i}|\rho_0\rangle\rangle$. After sufficient cycles, the decay mode of the survival probability is determined by the SDM. For the SDM with a real and negative eigenvalue, we have $P(m)\approx P_s+d_{s+1} e^{-\Gamma m (\tau+\tau_A)}$ with $P_s=\sum_{i=1}^s d_i$, then the envelope exponentially damps to $P_s$; For the SDM with a complex conjugate pair of eigenvalues, the survival probability can show damped oscillations. Note that as the invariant subspace is the maximally mixed state $\I/2$, so $P_s=\Tr(\rho_0)/2=1/2$.

For the closed qubit model, the exact survival probabilities across discrete measurement steps [Eq.~(\ref{eq:pm_prob})] agree well with the SDM predictions for various $\tau$ and $\tau_A$ [Figs.~\hyperref[fig:closesystem]{\ref*{fig:closesystem}(d)}--\hyperref[fig:closesystem]{\ref*{fig:closesystem}(f)}], confirming the validity of using the SDM to quantify the dynamics.
This response transition also has a simple physical interpretation. Starting from $\rho_0=\ket{+}\bra{+}$ with $|\pm\rangle$ being eigenstates of $\sigma_x$, the Hamiltonian $H=\omega\sigma_z/2$ rotates the qubit on the equator of the Bloch sphere, while the Ramsey measurement acts as a nonselective measurement in the $\sigma_x$ basis and removes coherence between $\ket{+}$ and $\ket{-}$. For a projective measurement with $g\tau_A=\pi/2$, the qubit evolves into an equal superposition of the two $\sigma_x$ eigenstates at time $\tau=\pi/(2\omega)$ and can decay to the maximally mixed state after a single measurement [Fig.~\hyperref[fig:closesystem]{\ref*{fig:closesystem}(f)}]. For weaker measurements with $g\tau_A<\pi/2$, a single measurement only partially suppresses the $\sigma_x$-basis coherence, so the fastest mixing occurs at times determined by $\tan^4(g\tau_A/2)=\sin^2(\omega\tau)$ along the exceptional line, where coherent rotation and partial measurement-induced dephasing cooperate most efficiently. This less intuitive weak-measurement case is precisely where the SDM criterion predicts the observed change of the survival-probability envelope in Figs.~\hyperref[fig:closesystem]{\ref*{fig:closesystem}(d)} and {\color{blue} {\ref*{fig:closesystem}(e)}}.

\begin{figure}[htbp]
    \centering
    \includegraphics[width=\linewidth]{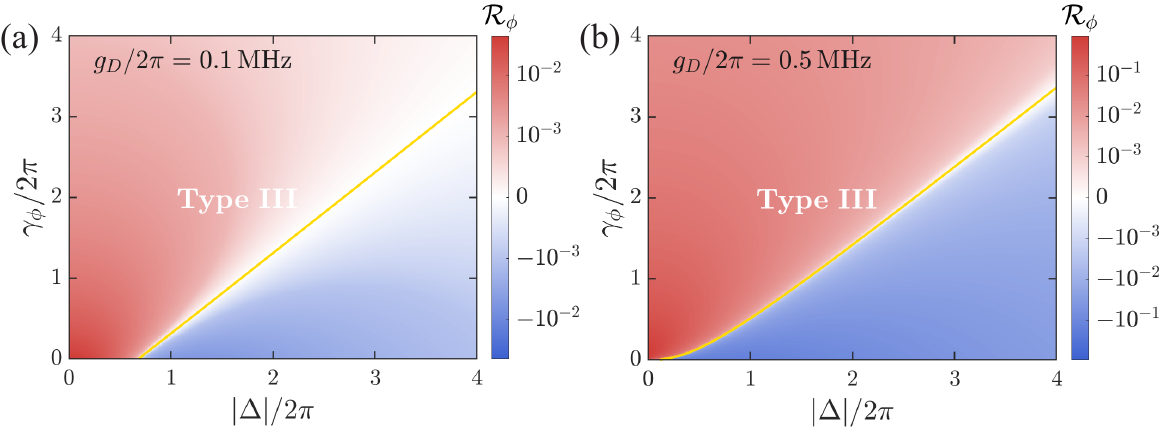}
    \caption{Quantum Zeno and anti-Zeno responses in continuous-time measurements. Here we consider a superconducting qubit under varying qubit-defect frequency detuning $\Delta$ and qubit measuring (dephasing) rate $\gamma_{\phi}$ with (a) a weak qubit-defect coupling ($g_{D}/2\pi=0.1$ MHz) and (b) a stronger one ($g_{D}/2\pi=0.5$ MHz). }

    \label{fig:opensystem}
\end{figure}

\prlsection{Extension to continuous-time measurement models}
Our framework can be directly extended to continuous-time measurement models, where the measurement acts as a dissipative process with the dissipation strength denoting the measurement rate \cite{Li2023,Greenfield2025,Thorbeck2024,Blumenthal2022}. Here we analyze a superconducting qubit-defect model introduced in Ref.~\cite{Thorbeck2024} to describe the readout-induced change of the qubit lifetime, where the Kofman--Kurizki spectral-overlap formula \cite{Kofman2000,Kofman2001} was derived by the dissipation incorporated self-consistent coarse-graining (DISCo) approach \cite{Thorbeck2024}. Our spectral formulation provides a more general interpretation from the Liouvillian gap, allowing the Zeno-to-anti-Zeno boundary to be identified from the derivative $\mathcal{R}_{\phi}=-\partial\Gamma/\partial\gamma_{\phi}$. By calculating the Liouvillian spectral gap within the single-excitation subspace, we can exactly reproduce the decay rate formula and predict the Zeno-to-anti-Zeno transitions beyond the weak qubit-defect coupling regime.


We consider a superconducting qubit coupled to a defect with the Hamiltonian under the rotating-wave approximation being $H_{qD} = \frac{\omega_{q}}{2}\sigma_{z}+\frac{\omega_{D}}{2}\tau_{z}+g_{D}(\sigma_{+}\eta_{-}+\sigma_{-}\eta_{+})$, where $\sigma_{i}$ ($\eta_{i}$) ($i=z,\pm$) denotes the Pauli operators of the qubit (defect), and $\omega_{q}$ ($\omega_{D}$) represents the energy splitting of the qubit (defect). Then the generator for the free evolution is $\mathcal{G}=-i[H_{qD},\cdot]+\gamma_{q}\mathcal{D}[\sigma_{-}]+\gamma_{D}\mathcal{D}[\eta_{-}]$, where $\gamma_{q}$ and $\gamma_{D}$ denote their corresponding energy relaxation rates. The continuous measurement can be modeled as a phase damping channel given by $\mathcal{G}_{m}=\frac{\gamma_{\phi}}{2}\mathcal{D}[\sigma_{z}]$, where $\gamma_{\phi}$ is the dephasing rate. So the effective generator $\mathcal{G}_{\mathrm{eff}}=\mathcal{G}+\mathcal{G}_{m}$ (see Sec. S3A of SM).

In the weak qubit-defect coupling regime [$g_{D}/\sqrt{a^{2}+\Delta^{2}}\ll 1$ with $a=\gamma_{\phi}+(\gamma_{D}-\gamma_{q})/2$], we can employ perturbation theory to evaluate the spectrum of $\mathcal{G}_{\mathrm{eff}}$. In the uncoupled limit ($g_{D}=0$), the four eigenvalues are given by $\lambda_{1}=-\gamma_{q}$, $\lambda_{2}=-\gamma_{D}$, $\lambda_{3}=-A+i\Delta$, and $\lambda_{4}=-A-i\Delta$, where $A = \gamma_{\phi} + \frac{\gamma_q + \gamma_{D}}{2}$ and $\Delta=\omega_{q}-\omega_{D}$. For $\gamma_{q}<\gamma_{D}$, $\lambda_{1}=-\gamma_{q}$ serves as the SDM. While for a weak $g_D$,  the perturbed eigenvalue can be calculated as $\lambda_{1}=-\gamma_{q}-2g_{D}^{2}\frac{a}{a^{2}+\Delta^{2}}$ (see Sec. S3B of SM).
Since the perturbative correction to the spectrum is of order $\mathcal{O}(g_{D}^{2})$, this mode retains its status as the SDM even after the perturbation, and the spectral gap is determined as $\Gamma=-\mathrm{Re}(\lambda_{\mathrm{SDM}})=\gamma_{q}+2g_{D}^{2}\frac{a}{a^{2}+\Delta^{2}}$. 

Then we can derive the response as $\mathcal{R}_{\phi}=-\pdv{\Gamma}{\gamma_{\phi}}=2 g_{D}^{2}\frac{a^{2}-\Delta^{2}}{(a^{2}+\Delta^{2})^{2}}$. We find the critical dephasing strength is $\gamma_{\phi}^{*} = |\omega_q - \omega_D| - \frac{\gamma_{D}}{2} + \frac{\gamma_q}{2}$, corresponding to a smooth extremum of the SDM [see the yellow line in Fig.~\hyperref[fig:opensystem]{\ref*{fig:opensystem}(a)}]. The system exhibits the Zeno (anti-Zeno) response when $\gamma_{\phi}>\gamma_{\phi}^{*}$ ($\gamma_{\phi}<\gamma_{\phi}^{*}$). 
Let us examine two specific scenarios. Under the resonance condition $\Delta=0$, the critical dephasing strength evaluates to $\gamma_{\phi}^{*}=\gamma_{q}/2-\gamma_{D}/2<0$, and the system exhibits a Zeno response (this conclusion also holds for small detuning satisfying $|\Delta|<\gamma_{D}/2-\gamma_{q}/2$). In contrast, for a large detuning satisfying $|\Delta|>\gamma_{D}/2-\gamma_{q}/2$, increasing $\gamma_{\phi}$ across the critical threshold $\gamma_{\phi}^{*}$ drives the system from anti-Zeno to Zeno response. As shown in Fig.~\hyperref[fig:opensystem]{\ref*{fig:opensystem}(b)}, increasing $g_{D}$ beyond the weak-coupling regime shifts the crossover boundary. At even larger $g_{D}$, this boundary asymptotically approaches a unity-slope line through the origin (see Sec. S3C in SM). This reveals that the Zeno-to-anti-Zeno transition in the strong-coupling regime is fundamentally governed by the competition between the detuning $|\Delta|$ and dephasing $\gamma_{\phi}$.

\prlsection{Conclusions and outlooks}
We have developed a general framework for characterizing quantum Zeno and anti-Zeno responses of a generic quantum system under repetitive measurements of arbitrary strength. By describing each evolution--measurement cycle as a quantum channel and defining an effective Liouvillian generator, we distinguish the quantum Zeno and anti-Zeno responses according to how the corresponding spectral gap varies with the measurement frequency. Through an exactly solvable qubit model under discrete-time measurements, we show that the transitions between Zeno and anti-Zeno responses can arise from spectral coalescence at EPs or spectral crossings of competing decay modes, with clear observable signatures of the evolution of the survival probability . Such a transition can also occur in continuous-time measurement models with smooth extrema of the spectral gap.

The framework can be extended to more complex evolution and measurement in generic quantum systems, such as quantum non-Markovian dynamics under structured environments \cite{Vega2017,Breuer2016,Tamascelli2018}, quantum feedback-assisted dynamics  \cite{Wiseman1993,Wiseman1994,Nakagawa2025,Vijay2012}, and many-body systems with multiple competing slow modes \cite{Haga2024,Rose2016}. We also expect more discoveries to be made about the fundamental relation between the structures of evolution-measurement channels and the Zeno-to-anti-Zeno response transition, and investigate how symmetry and topology of the channel constrain the response phase diagrams \cite{Nakagawa2025}. Additionally, it will be interesting to explore similar spectral criteria to quantify the dynamics of quantum correlations \cite{Francica2010}, Floquet Liouvillian engineering \cite{Chen2025}, many-body entanglement transition \cite{Li2018,Choi2020,Mochizuki2025,Oshima2025} and deep thermalization \cite{Feng2026}.


The research is supported by the National Natural Science Foundation of China (No. 12574082, No. E31Q02BG), the Chinese Academy of Sciences (No. E0SEBB11, No. E27RBB11), Quantum Science and Technology-National Science and Technology Major Project (No. 2021ZD0302300) and Chinese Academy of Sciences Project for Young Scientists in Basic Research (YSBR-090).

\bibliography{zeno.bib}

\section*{End Matter}
\prlsection{Correspondence between discrete- and continuous-time measurement models}
The connection between the discrete- and continuous-time models stems from the insight that the former reduces to the latter in the limit of vanishing measurement and evolution times. 

Within our framework, the discrete-time channel model consists of a free-evolution channel of duration $\tau$ and a measurement channel of duration $\tau_{A}$. These two generally noncommuting operations yield the generic expression $\Phi(\tau,\tau_{A})=\mathcal{M}_{\tau_{A}}e^{\mathcal{G}\tau}$. For weak and short-duration measurements, the measurement operation can be effectively treated as being driven by a measurement generator, yielding $\mathcal{M}_{\tau_{A}}=e^{\mathcal{G}_{m}\tau_{A}}$ with $\mathcal{G}_{m}=\gamma_\phi^{\mathrm{eff}} \mathcal{D}$. Employing the Lie-Trotter product formula \cite{Trotter1959,Han2021},
\begin{align}
    \Phi^m(\tau,\tau_{A})&=(e^{\mathcal{G}_{m}\tau_{A}}e^{\mathcal{G}\tau})^{m}\\ \notag
    &=e^{m(\mathcal{G}\tau+\mathcal{G}_{m}\tau_{A})}+\mathcal{O}\!\left(m\tau\tau_A\left\|[\mathcal{G}_m,\mathcal{G}]\right\|\right),
\end{align}
For the total time $t=m(\tau+\tau_{A})$, the effective generator reads ${\mathcal{G}}_{\mathrm{eff}} = (\tau{\mathcal{G}} + \tau_{A}{\mathcal{G}}_{m}) / (\tau+\tau_{A})$, where $\tau/(\tau+\tau_{A})$ and $\tau_{A}/(\tau+\tau_{A})$ serve as the mixing weights for the two generators.

In the discrete-time model, the response is defined as $\mathcal{R}(\tau,\tau_{A})=-\pdv{\Gamma}{f}$, with $f=1/(\tau+\tau_{A})$. In the continuous-time limit, the absence of explicit intervals $\tau$ and $\tau_{A}$ renders the measurement frequency $f$ ill-defined, necessitating its replacement with an alternative physical parameter. We first rewrite the discrete-time channel as
\begin{align}
    {\Phi}(\tau,\tau_{A})=e^{\mathcal{G_{\rm{eff}}}(\tau+\tau_{A})}=e^{r(\mathcal{G}+\frac{\tau_{A}}{\tau}\gamma_\phi^{\mathrm{eff}} \mathcal{D})(\tau+\tau_{A})},
\end{align}
with $r=\tau/(\tau+\tau_{A})$. Since the general form of the channel generator derived via the master equation is $\mathcal{G}+\gamma_{\phi}\mathcal{D}$, matching the coefficients yields $\gamma_{\phi}=\frac{\tau_{A}}{\tau}\gamma_{\phi}^{\rm{eff}}=\frac{\tau_A\gamma_{\phi}^{\rm{eff}}}{r} f\propto f$. According to the chain rule for derivatives, the continuous-time response defined by $\mathcal{R}_{\phi}=-\pdv{\Gamma}{\gamma_{\phi}}$ differs from the discrete-time one only by a scaling factor $\frac{\tau_{A}\gamma_{\phi}^{\mathrm{eff}}}{r^{2}}$. 

\prlsection{Ramsey measurement as an effective dissipative generator}
The Ramsey-interferometry measurement used in the closed-qubit model provides a concrete realization of the measurement generator assumed above. The ancilla qubit is initialized in $\ket{0}_{q}$, rotated by $R_{\phi_1}$, coupled to the measured system through $H_A=\sigma_q^z\otimes A$ for a time $\tau_A$, rotated by $R_{\phi_2}$, and finally measured in the $\sigma_q^z$ basis. For the outcome $\alpha=0,1$, the resulting Kraus operator is
\begin{equation}
M_\alpha=\frac{1}{2}\left[U_+-(-1)^\alpha e^{i\Delta\phi}U_-\right],\quad
U_\pm=e^{\mp iA\tau_A},
\end{equation}
where $\Delta\phi=\phi_1-\phi_2$ \cite{Pfender2019,Jin2026}. For the nonselective measurement channel, the outcome-dependent phase drops out after summing over $\alpha$, giving
\begin{equation}
\mathcal{M}_{\tau_A}(\rho)=\sum_{\alpha=0,1}M_\alpha\rho M_\alpha^\dagger
=\frac{1}{2}\left(U_+\rho U_+^\dagger+U_-\rho U_-^\dagger\right).
\end{equation}
For the qubit example considered in the main text, $A=g\sigma_x/2$. With $\mu=g\tau_A$, the vectorized measurement channel in the basis $\{\ket{0}\bra{0},\ket{0}\bra{1},\ket{1}\bra{0},\ket{1}\bra{1}\}$ reads
\begin{equation}
\hat{\mathcal{M}}_{\tau_A}=
\begin{pmatrix}
\cos^2\frac{\mu}{2} & 0 & 0 & \sin^2\frac{\mu}{2}\\
0 & \cos^2\frac{\mu}{2} & \sin^2\frac{\mu}{2} & 0\\
0 & \sin^2\frac{\mu}{2} & \cos^2\frac{\mu}{2} & 0\\
\sin^2\frac{\mu}{2} & 0 & 0 & \cos^2\frac{\mu}{2}
\end{pmatrix}.
\end{equation}
In the weak-measurement limit $g\tau_A\ll1$, this channel becomes \cite{Jin2026}
\begin{equation}
\hat{\mathcal{M}}_{\tau_A}\simeq
\mathbb{I}+\frac{g^2\tau_A^2}{4}
\left(\sigma_x\otimes\sigma_x-\mathbb{I}\right)
\simeq e^{\hat{\mathcal{G}}_m\tau_A},
\end{equation}
with
\begin{equation}
\hat{\mathcal{G}}_m=\gamma_\phi^{\rm eff}\hat{\mathcal{D}}[\sigma_x],
\quad
\gamma_\phi^{\rm eff}=\frac{g^2\tau_A}{4}.
\end{equation}
Thus a short-time Ramsey weak measurement is equivalent to a dissipative generator that dephases the measured system in the $\sigma_x$ basis, providing the explicit measurement-level origin of the discrete-to-continuous correspondence.




\end{document}


\title{ Supplemental Material for "Quantum Zeno and Anti-Zeno Responses: Universal Spectral Criterion for Measurement-Induced Decay"}

\author{Yu-Xiang Chen  }
\thanks{These authors contributed equally to this work.}
\affiliation{State Key Laboratory of Semiconductor Physics and Chip Technologies, Institute of Semiconductors, Chinese Academy of Sciences, Beijing 100083, China}
\affiliation{Center of Materials Science and Opto-Electronic Technology, University of Chinese Academy of Sciences, Beijing 100049, China}

\author{Yuan-De Jin }
\thanks{These authors contributed equally to this work.}
\affiliation{State Key Laboratory of Semiconductor Physics and Chip Technologies, Institute of Semiconductors, Chinese Academy of Sciences, Beijing 100083, China}
\affiliation{Center of Materials Science and Opto-Electronic Technology, University of Chinese Academy of Sciences, Beijing 100049, China}

\author{Wen-Long Ma}
\email{wenlongma@semi.ac.cn}
\affiliation{State Key Laboratory of Semiconductor Physics and Chip Technologies, Institute of Semiconductors, Chinese Academy of Sciences, Beijing 100083, China}
\affiliation{Center of Materials Science and Opto-Electronic Technology, University of Chinese Academy of Sciences, Beijing 100049, China}

\begin{abstract}

\end{abstract}

\maketitle
\tableofcontents

\section{Measurement-induced decay for discrete-time measurement models}
\subsection{Diagonalizable channels}
If the quantum channel is diagonalizable, the quantum channel for each evolution-measurement cycle can be spectrally decomposed as 
\begin{equation}
\hat\Phi(\tau,\tau_A)=\mathcal{M}_{\tau_{A}} e^{\mathcal{G}\tau}=\sum\limits_{i=1}^{d^{2}}\Lambda_{i}\hket{R_{i}}\hbra{L_{i}},
\end{equation}
where $\Lambda_i$ denotes the eigenvalue of the quantum channel, and $\{\hket{R_i},\hket{L_i}\}$ is a complete biorthonormal system satisfying $\brakett{L_i}{R_j}=\delta_{ij}$. Then the quantum channel can be regarded as being induced by an effective generator $e^{\hat{\mathcal{G}}_{\rm eff}}$ for a duration $\tau+\tau_A$,
 \begin{equation}
     \phii=e^{\hat{\mathcal{G}}_{\rm eff}(\tau+\tau_A)}=\sum\limits_{i=1}^{d^{2}}e^{\lambda_{i}(\tau+\tau_{A})}\hket{R_{i}}\hbra{L_{i}},
 \end{equation}
with the eigenvalue of the generator $\hat{\mathcal{G}}_{\rm eff}$ being $\lambda_i=(\ln\Lambda_i)/(\tau+\tau_A)$.
 

 As $\hat\Phi(\tau,\tau_A)$ is a completely positive and trace-preserving (CPTP) map, its spectrum $\{\Lambda_i\}$ is confined to the unit disk in the complex plane \cite{Kraus1983,Watrous2018}. Consequently, the eigenvalue $\lambda_{i}$ of $\hat{\mathcal{G}}_{\rm eff}$ satisfies $\mathrm{Re}(\lambda_{i})\leq 0$ and $\mathrm{Im}(\lambda_{i}) \in [0,2\pi/(\tau+\tau_A)]$. To facilitate a straightforward comparison within the decay spectrum, we impose the ordering relation $\mathrm{Re}(\lambda_{i+1})\leq \mathrm{Re}(\lambda_{i})\leq 0$. Since the complex eigenvalues of a quantum channel appear in conjugate pairs \cite{Jin2024}, this property carries over to the decay spectrum: if $\mathrm{Im}(\lambda_{i})\neq 0$, there exists a $\lambda_{i+1}$ such that $\mathrm{Re}(\lambda_{i})=\mathrm{Re}(\lambda_{i+1})$ and $\mathrm{Im}(\lambda_{i+1})=-\mathrm{Im}(\lambda_{i})$. We define the slowest decay mode (SDM) as the spectral component with the largest non-zero real part, namely $\lambda_{s+1}$, and refer to $|\mathrm{Re}(\lambda_{s+1})|$ as the Liouvillian gap \cite{Macieszczak2016,Minganti2018} ($s$ is the number of fixed points). In particular, if $\mathrm{Re}(\lambda_{s+1})=\mathrm{Re}(\lambda_{s+k})$ and $\mathrm{Re}(\lambda_{s+k+1})<\mathrm{Re}(\lambda_{s+k})$, the set of modes $\{ \lambda_{i} \}_{i=s+1}^{sk}$ is collectively referred to as the SDMs. Apart from the SDMs, all other spectral components with non-zero real parts are collectively referred to as the fast decay modes (FDMs).

 In what follows, we will elucidate how the SDM can be utilized to calculate the system's decay rate and justify why the SDM serves as an accurate characterization of the overall decay dynamics. Assuming $m$ successive applications of the channel on an initial state $\hhket{\rho}$, we have
\begin{equation}
\hat{\Phi}^{m}\hhket{\rho}=\sum_{i=1}^{d^{2}}c_{i} e^{m \lambda_{i}(\tau+\tau_{A})}\hhket{R_{i}}.
\end{equation}
 with $c_{i}=\langle \langle L_{i} | \rho \rangle \rangle$. Focusing on a specific term $c_{k}e^{m\lambda_{k}(\tau+\tau_{A})}\hhket{R_{k}}$, we assume without loss of generality that $\lambda_{k}$ is complex with a nonvanishing imaginary part. This necessitates a conjugate spectra of generator $\lambda_{k+1}=\lambda_{k}^{*}$ with the corresponding coefficient $c_{k+1}=c_{k}^{*}$. Assuming $c_{k}=ce^{i\delta}$ and $c_{k+1}=ce^{-i\delta}$. Summing this conjugate pair of decay terms yields
 \begin{align}
     &c_{k}e^{m\lambda_{k}(\tau+\tau_{A})}\hhket{R_{k}}+c_{k+1}e^{m\lambda_{k+1}(\tau+\tau_{A})}\hhket{R_{k+1}} \\ \notag
     = &ce^{\mathrm{Re}(\lambda_{k})t}(e^{i[m(\tau+\tau_{A})\mathrm{Im}(\lambda_{k})+\delta]}\hhket{R_{k}}+e^{-i[m(\tau+\tau_{A})\mathrm{Im}(\lambda_{k})+\delta]}\hhket{R_{k+1}})\\ \notag
     =&ce^{\mathrm{Re}(\lambda_{k})t}[\cos(\psi)\hhket{R^{+}}+i\sin(\psi)\hhket{R^{-}}],
 \end{align}
 where $\hhket{R^{\pm}}=\hhket{R_{k}}\pm\hhket{R_{k+1}}$, and $\psi=m(\tau+\tau_{A})\mathrm{Im}(\lambda_{k})+\delta$ is an $m$-dependent phase. Since $t=m(\tau+\tau_{A})$ represents the total evolution time of the system, $-\mathrm{Re}(\lambda_{k})$ characterizes the decay rate of this conjugate pair of decay terms, while $\mathrm{Im}(\lambda_{k})$ solely determines the oscillation frequency. Each generator spectrum $\lambda_{k}$ corresponds to a decay rate $\gamma_{k}=-\mathrm{Re}(\lambda_{k})$.

 Intuitively, faster-decaying terms vanish rapidly before the slower ones exhibit any discernible decay, whereas slower-decaying terms persist long after the others have diminished to negligible levels. For a specific mode $\lambda_k$, assumed to be real without loss of generality (a complex eigenvalue is simply treated as a conjugate pair), we can identify the time window $(1/\gamma_{k+1}, 1/\gamma_{k})$ as its characteristic interval, during which $\lambda_k$ dominates the system's decay. This inherently assumes a finite gap between $\mathrm{Re}(\lambda_k)$ and $\mathrm{Re}(\lambda_{k+1})$; otherwise, if $\mathrm{Re}(\lambda_k) \approx \mathrm{Re}(\lambda_{k+1})$, these terms would simply merge into a single decay mode with an aggregated coefficient $c_k + c_{k+1}$. In particular, the SDM $\lambda_{\mathrm{SDM}}$ dominates the decay not only within the interval $(1/\gamma_{2}, 1/\gamma_{\mathrm{SDM}})$, but also in the regime $(1/\gamma_{\mathrm{SDM}}, +\infty)$ since no slower decay modes exist. Consequently, when considering the long-time decay (i.e., for large $m$), the action of the channel can be simplified to
\begin{align}
\hat{\Phi}^{m}\hhket{\rho} &\simeq \sum_{i=1}^{s}c_{i}\hhket{R_{i}}+c_{s+1}e^{m[\mathrm{Re}(\lambda_{\mathrm{SDM}})](\tau + \tau_{A})}\hhket{R_{s+1}}\\ \notag 
&=\sum_{i=1}^{s}c_{i}\hhket{R_{i}}+c_{s+1}e^{-m\Gamma(\tau+\tau_{A})}\hhket{R_{s+1}}.
\end{align}
However, if $c_{s+1}=0$, the contribution from the second decay term becomes non-negligible. Under this specific circumstance, the long-time decay is governed by the second SDM. Such an initial-state-dependent property of the channel decay provides a natural explanation for the quantum Mpemba effect.
 
\subsection{Non-diagonalizable channels}
The above discussion can be extended to non-diagonalizable channels. A non-diagonalizable quantum channel admits the Jordan decomposition
\begin{equation}
\begin{aligned}
\hat{\Phi}
&=S\left(\bigoplus_{i=1}^{\kappa}\mathcal{J}_{d_i}(\Lambda_i)\right)S^{-1}\\
&=S\left[
\sum_{|\Lambda_j|=1}\Lambda_j\mathcal{P}_j
+\sum_{|\Lambda_i|<1}\left(\Lambda_i\mathcal{P}_i+\mathcal{N}_i\right)
\right]S^{-1}.
\end{aligned}
\end{equation}
Here, $i$ labels an individual Jordan block, $d_i$ is its dimension, and $\mathcal{P}_i$ is the projector onto that block in the Jordan basis, satisfying $\mathcal{P}_i\mathcal{P}_j=\delta_{ij}\mathcal{P}_i$. The nilpotent part $\mathcal{N}_i$ satisfies $\mathcal{P}_i\mathcal{N}_i=\mathcal{N}_i\mathcal{P}_i=\mathcal{N}_i$ and $\mathcal{N}_i^{d_i}=0$. For $d_i=1$, the block is trivial and $\mathcal{N}_i=0$. The peripheral blocks with $|\Lambda_j|=1$ are semisimple, as required by the bounded powers of a finite-dimensional quantum channel, whereas nontrivial Jordan blocks may occur in the decaying sector $|\Lambda_i|<1$.

Because $\Lambda_i\mathcal{P}_i$ commutes with $\mathcal{N}_i$, the binomial theorem gives
\begin{equation}
\left(\Lambda_i\mathcal{P}_i+\mathcal{N}_i\right)^m
=
\sum_{\ell=0}^{d_i-1}
\binom{m}{\ell}
\Lambda_i^{m-\ell}
\mathcal{N}_i^{\ell}\mathcal{P}_i.
\end{equation}
The sum terminates because $\mathcal{N}_i^{d_i}=0$. In the long-time regime considered below, $m\gg\max_i(d_i)$, so that the upper limit can be written directly as $d_i-1$. It follows that
\begin{equation}
\hat{\Phi}^{m}
=S\left[
\sum_{|\Lambda_j|=1}\Lambda_j^m\mathcal{P}_j
+\sum_{|\Lambda_i|<1}\sum_{\ell=0}^{d_i-1}
\binom{m}{\ell}\Lambda_i^{m-\ell}
\mathcal{N}_i^{\ell}\mathcal{P}_i
\right]S^{-1}.
\end{equation}

Given an initial state $\hhket{\rho}$, we define $\hhket{\rho_{i,\ell}}=S\mathcal{N}_i^{\ell}\mathcal{P}_iS^{-1}\hhket{\rho}$.
Then $m$ iterations of the channel give
\begin{equation}
\hat{\Phi}^{m}\hhket{\rho}=\sum_{|\Lambda_j|=1}\Lambda_j^m\hhket{\rho_{j,0}}+\sum_{|\Lambda_i|<1}\sum_{\ell=0}^{d_i-1}\binom{m}{\ell}\Lambda_i^{m-\ell}\hhket{\rho_{i,\ell}}.
\end{equation}
Thus, a nontrivial Jordan block generally produces several generalized-eigenvector components and cannot be represented by a universal scalar polynomial multiplying a single fixed state.

A scalar polynomial prefactor can instead be defined after projecting the state onto a linear observable. For a decay signal $F(m)=\langle\!\langle O|\hat{\Phi}^{m}|\rho\rangle\!\rangle$, which reduces to the survival probability $P(m)$ for $O=\rho=\rho_0$, let $a_{i\ell}=\langle\!\langle O|\rho_{i,\ell}\rangle\!\rangle$. The contribution from a decaying Jordan block with $\Lambda_i\neq0$ is
\begin{equation}
F_i(m)=\sum_{\ell=0}^{d_i-1}a_{i\ell}\binom{m}{\ell}\Lambda_i^{m-\ell}=c_i(m)\Lambda_i^m.
\end{equation}
where $c_i(m)=\sum_{\ell=0}^{d_i-1}a_{i\ell}\binom{m}{\ell}\Lambda_i^{-\ell}$.
The coefficient $c_i(m)$ is a polynomial in $m$ of degree at most $d_i-1$, with coefficients determined by the initial state, the observable, and the generalized eigenvectors of the Jordan block.

Writing $\Lambda_i=e^{\lambda_i(\tau+\tau_A)}$ and $\Gamma_i=-\mathrm{Re}(\lambda_i)$, the magnitude of a nonzero block contribution is
\begin{equation}
|F_i(m)|
=|c_i(m)|e^{-m\Gamma_i(\tau+\tau_A)}
=e^{-m\Gamma_i^{\mathrm{eff}}(m)(\tau+\tau_A)},
\end{equation}
with the finite-time effective decay rate $\Gamma_i^{\mathrm{eff}}(m)=\Gamma_i-R_i(m)$, where $R_i(m)=\ln|c_i(m)|/[m(\tau+\tau_A)]$.
If the highest nonvanishing Jordan contribution has order $p_i\leq d_i-1$, then $c_i(m)\sim C_i m^{p_i}$ and
\begin{equation}
R_i(m)
\sim\frac{p_i\ln m+\ln|C_i|}{m(\tau+\tau_A)}
\longrightarrow0
\qquad(m\to\infty).
\end{equation}
Therefore, the Jordan polynomial modifies the finite-time decay profile but does not change the asymptotic decay rate $\Gamma_i$. For comparison with a decay curve over a finite observation window, the SDM correction may be evaluated at a characteristic cycle number $m_0=C/[\Gamma_{\mathrm{SDM}}(\tau+\tau_A)]$. A smaller $C$ probes the earlier decay stage, whereas a larger $C$ probes the later-time regime. This characteristic-cycle prescription is a finite-time fitting convention rather than a condition for the Jordan expansion itself.

\section{Details for discrete-time measurement models}
\subsection{Closed qubit model}

\subsubsection{Spectra of the channel and generator}
\begin{figure}[H]
\centering
  \includegraphics[width=15cm]{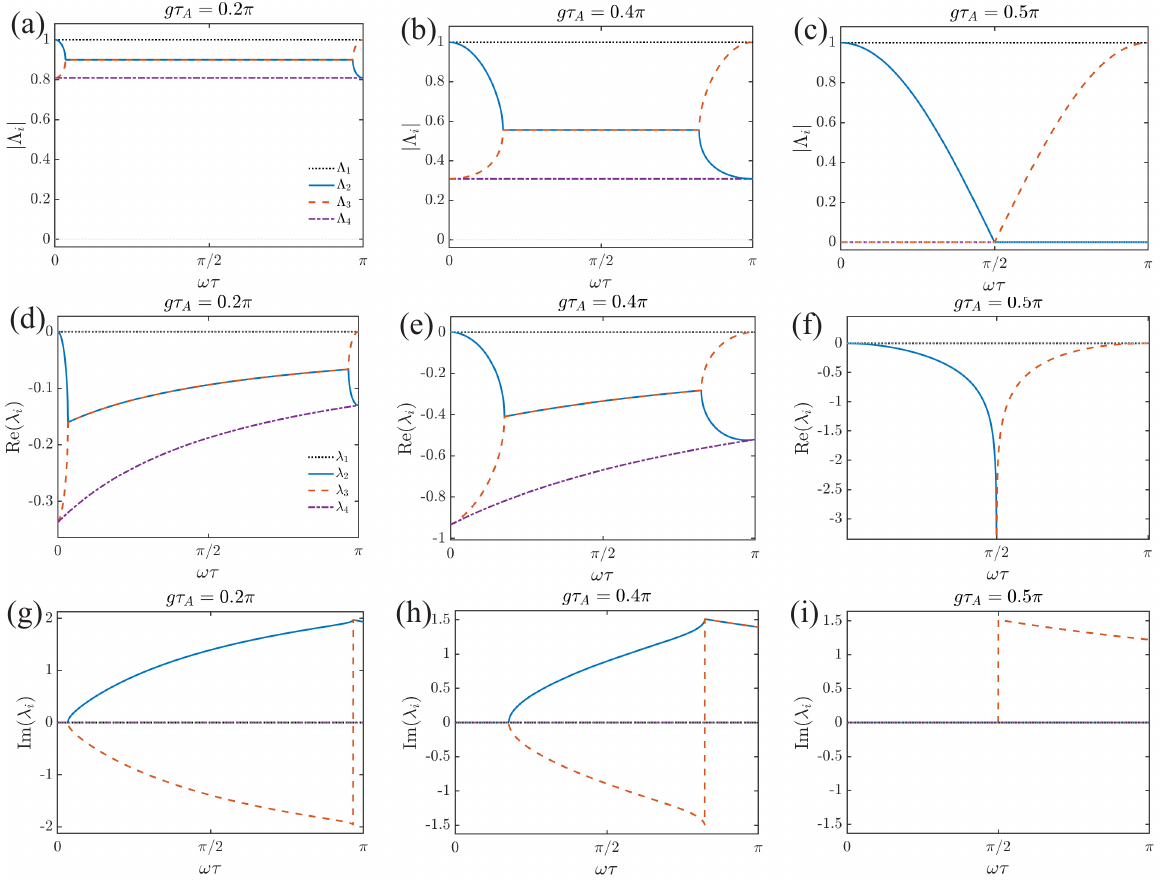}
  \caption{Spectrum of the closed-qubit  channel.
Columns correspond to $g\tau_A=0.2\pi$, $0.4\pi$, and $0.5\pi$, while rows show $|\Lambda_i|$, $\mathrm{Re}(\lambda_i)$, and $\mathrm{Im}(\lambda_i)$ as functions of $\omega\tau$, with generator spectrum $\lambda_i=\ln\Lambda_i/(\tau+\tau_A)$.}
  \label{close_spectrum}
\end{figure}

We first derive the exact spectrum of the one-cycle channel for the closed-qubit model. The free evolution is generated by $H=\omega\sigma_z/2$, and the Ramsey measurement acts as a nonselective measurement along the $x$ direction. For compactness, we denote $\theta=\omega\tau$ and $\beta=\cos(g\tau_A)$. For the Ramsey measurement used in the main text, the two measurement outcomes are described by the Kraus operators
\begin{equation}
M_{\alpha}=\frac{1}{\sqrt{2}}\left(c_{+}\mathbb{I}+(-1)^{\alpha}c_{-}\sigma_x\right),
\qquad
c_{\pm}=\sqrt{\frac{1\pm \cos(g\tau_A)}{2}},
\end{equation}
with $\alpha=0,1$. The nonselective measurement channel $\mathcal{M}_{\tau_A}(\rho)=\sum_{\alpha}M_{\alpha}\rho M_{\alpha}^{\dagger}$ is therefore a tunable dephasing channel in the $\sigma_x$ basis. Since the Kraus operators are built from $\mathbb{I}$ and $\sigma_x$, the measured-axis component $\sigma_x$ is unchanged, whereas the two transverse components are suppressed by $c_+^2-c_-^2=\cos(g\tau_A)=\beta$. Thus $\mathcal{M}_{\tau_A}(\mathbb{I})=\mathbb{I}$, $\mathcal{M}_{\tau_A}(\sigma_x)=\sigma_x$, and $\mathcal{M}_{\tau_A}(\sigma_{y,z})=\beta\sigma_{y,z}$. Combining this measurement action with the $z$-axis rotation generated by $H$, one complete evolution-measurement cycle acts as
\begin{equation}
\begin{aligned}
\mathbb{I} &\mapsto \mathbb{I},\\
\sigma_x &\mapsto \cos\theta\,\sigma_x+\beta\sin\theta\,\sigma_y,\\
\sigma_y &\mapsto -\sin\theta\,\sigma_x+\beta\cos\theta\,\sigma_y,\\
\sigma_z &\mapsto \beta\,\sigma_z .
\end{aligned}
\end{equation}
The identity component immediately gives the invariant channel eigenvalue $\Lambda_1=1$, and the $\sigma_z$ component gives $\Lambda_4=\beta=\cos(g\tau_A)$. The remaining two eigenvalues come from the two-dimensional block acting on $\{\sigma_x,\sigma_y\}$,
\begin{equation}
B_{xy}=
\begin{pmatrix}
\cos\theta & -\sin\theta\\
\beta\sin\theta & \beta\cos\theta
\end{pmatrix}.
\end{equation}
Solving $\det(B-\Lambda \mathbb{I})=0$ gives
\begin{equation}
\Lambda^2-(1+\beta)\cos\theta\,\Lambda+\beta=0,
\end{equation}
and hence
\begin{equation}
\Lambda_{2,3}
=
\frac{(1+\beta)\cos\theta
\pm
\sqrt{(1+\beta)^2\cos^2\theta-4\beta}}{2}.
\end{equation}
Using $\beta=\cos(g\tau_A)$ and $(1+\beta)/2=\cos^2(g\tau_A/2)$, the four eigenvalues of the one-cycle channel can be written as
\begin{equation}
\Lambda_1=1,\quad
\Lambda_{2,3}=
\cos^2\frac{g\tau_A}{2}
\left[
\cos(\omega\tau)
\pm
\sqrt{
\tan^4\frac{g\tau_A}{2}
-
\sin^2(\omega\tau)
}
\right],
\quad
\Lambda_4=\cos(g\tau_A).
\end{equation}
The spectrum of the generator is then defined by $\Lambda_i=\exp[\lambda_i(\tau+\tau_A)]$, namely
\begin{equation}
\lambda_i=\frac{\ln\Lambda_i}{\tau+\tau_A}.
\end{equation}
 The resulting channel-eigenvalue moduli and effective-generator spectra for representative measurement strengths are shown in Fig.~\hyperref[close_spectrum]{\ref*{close_spectrum}}. In particular, the pair $\Lambda_{2,3}$ coalesces when the square-root discriminant vanishes, giving the exceptional line $\tan^4(g\tau_A/2)=\sin^2(\omega\tau)$. This is the spectral condition underlying the exceptional-point response boundary discussed in the main text.

\subsubsection{Fixed points}
We next show that, in the nontrivial parameter regime considered in the main text, the maximally mixed state is the unique fixed point of the one-cycle channel. First, $\rho_{\rm ss}=\mathbb{I}/2$ is a fixed point: the unitary part satisfies $U(\tau)(\mathbb{I}/2)U^{\dagger}(\tau)=\mathbb{I}/2$, and the Ramsey measurement is unital because the terms proportional to $(-1)^\alpha\sigma_x$ cancel after summing over the two outcomes,
\begin{equation}
\sum_{\alpha=0,1}M_{\alpha}M_{\alpha}^{\dagger}
=
(c_+^2+c_-^2)\mathbb{I}
=
\mathbb{I}.
\end{equation}
Thus $\Phi(\tau,\tau_A)(\mathbb{I}/2)=\mathbb{I}/2$. To prove uniqueness, let a general qubit state be written as $\rho=(\mathbb{I}+r_x\sigma_x+r_y\sigma_y+r_z\sigma_z)/2$. Using the channel action derived above, the fixed-point condition $\Phi(\rho)=\rho$ gives
\begin{equation}
\begin{aligned}
r_x&=\cos\theta\,r_x-\sin\theta\,r_y,\\
r_y&=\beta\sin\theta\,r_x+\beta\cos\theta\,r_y,\\
r_z&=\beta r_z.
\end{aligned}
\end{equation}
The last equation gives $(1-\beta)r_z=0$. Excluding the zero-measurement case $\beta=1$, we have $r_z=0$. The remaining two equations can be written as
\begin{equation}
\begin{pmatrix}
1-\cos\theta & \sin\theta\\
-\beta\sin\theta & 1-\beta\cos\theta
\end{pmatrix}
\begin{pmatrix}
r_x\\
r_y
\end{pmatrix}
=0.
\end{equation}
A nonzero solution for $(r_x,r_y)$ exists only if the determinant of this coefficient matrix vanishes. Direct evaluation gives
$(1-\cos\theta)(1-\beta\cos\theta)+\beta\sin^2\theta
=(1+\beta)(1-\cos\theta)$.
Since $\beta\neq -1$ in the present Ramsey-measurement setting, and away from the trivial free-evolution points $\theta=2\pi n$ one has $1-\cos\theta\neq0$, the determinant is nonzero. Therefore $r_x=r_y=0$. Together with $r_z=0$, this implies that any fixed point must be $\rho_{\rm ss}=\mathbb{I}/2$. Hence the maximally mixed state is the unique fixed point of the closed-qubit channel in the parameter regime studied here.

\subsubsection{Zeno and anti-Zeno responses}
We now show explicitly why the region below the lower exceptional line gives a positive response for the closed-qubit model. We consider the parameter regime $0<g\tau_A<\pi/2$ and fix the measurement duration $\tau_A$, as in the definition of the discrete-time response in the main text. Below the lower exceptional line, one has
\begin{equation}
0<\omega\tau<\omega\tau_{\mathrm{EP}},
\qquad
\sin(\omega\tau_{\mathrm{EP}})
=
\tan^2\frac{g\tau_A}{2},
\end{equation}
and therefore
\begin{equation}
\tan^4\frac{g\tau_A}{2}-\sin^2(\omega\tau)>0 .
\end{equation}
In this region the two branches $\Lambda_{2,3}$ are real, and the SDM is the larger branch
\begin{equation}
\Lambda_2=
\cos^2\frac{g\tau_A}{2}
\left[
\cos(\omega\tau)
+
\sqrt{
\tan^4\frac{g\tau_A}{2}
-
\sin^2(\omega\tau)
}
\right].
\end{equation}
The corresponding decay rate is
\begin{equation}
\Gamma=-\frac{\ln\Lambda_2}{\tau+\tau_A},
\end{equation}
where $\Lambda_2=1$ at $\tau=0$. Since $g\tau_A$ is fixed, differentiating the logarithm of the above branch gives
\begin{equation}
\frac{\partial \ln\Lambda_2}{\partial\tau}
=
-\omega
\frac{\sin(\omega\tau)}
{
\sqrt{
\tan^4\frac{g\tau_A}{2}
-
\sin^2(\omega\tau)
}
}.
\end{equation}
The derivative is negative below the lower exceptional line because both $\sin(\omega\tau)$ and the square root in the denominator are positive. Using the response definition
\begin{equation}
\mathcal{R}
=
(\tau+\tau_A)^2
\frac{\partial\Gamma}{\partial\tau},
\end{equation}
we obtain
\begin{equation}
\mathcal{R}
=
\ln\Lambda_2
-
(\tau+\tau_A)
\frac{\partial\ln\Lambda_2}{\partial\tau}
=
\ln\Lambda_2
+
\omega(\tau+\tau_A)
\frac{\sin(\omega\tau)}
{
\sqrt{
\tan^4\frac{g\tau_A}{2}
-
\sin^2(\omega\tau)
}
}.
\end{equation}
It remains to show that the positive second term is always larger than $-\ln\Lambda_2$. From the derivative above and the initial value $\Lambda_2(0)=1$, we can write
\begin{equation}
-\ln\Lambda_2
=
\int_0^\tau d\tau'\,
\omega
\frac{\sin(\omega\tau')}
{
\sqrt{
\tan^4\frac{g\tau_A}{2}
-
\sin^2(\omega\tau')
}
}.
\end{equation}
The integrand is positive below the lower exceptional line. Moreover, it increases monotonically with $\tau'$, since
\begin{equation}
\frac{\partial}{\partial\tau'}
\left[
\omega
\frac{\sin(\omega\tau')}
{
\sqrt{
\tan^4\frac{g\tau_A}{2}
-
\sin^2(\omega\tau')
}
}
\right]
=
\omega^2
\frac{
\tan^4(g\tau_A/2)\cos(\omega\tau')
}{
\left[
\tan^4(g\tau_A/2)-\sin^2(\omega\tau')
\right]^{3/2}
}
>0 .
\end{equation}
Therefore, for any point below the lower exceptional line,
\begin{equation}
-\ln\Lambda_2
<
\tau\,
\omega
\frac{\sin(\omega\tau)}
{
\sqrt{
\tan^4\frac{g\tau_A}{2}
-
\sin^2(\omega\tau)
}
}
<
(\tau+\tau_A)
\omega
\frac{\sin(\omega\tau)}
{
\sqrt{
\tan^4\frac{g\tau_A}{2}
-
\sin^2(\omega\tau)
}
}.
\end{equation}
Substituting this inequality into the expression for $\mathcal{R}$ gives $\mathcal{R}>0$. Thus the real $\Lambda_2$ branch below the lower exceptional line indeed corresponds to the Zeno response region discussed in the main text.

\subsubsection{Physical picture for the response transition}
Since the fixed point of the closed-qubit channel is the maximally mixed state, the mixing rate can be understood as the rate at which the evolution--measurement cycle removes the Bloch-vector components of the initial state. We now make this picture explicit for the initial state used in the main text, $\rho_0=\ket{+}\bra{+}$, where $\ket{\pm}=(\ket{0}\pm\ket{1})/\sqrt{2}$ are the two eigenstates of $\sigma_x$. Under the free Hamiltonian $H=\omega\sigma_z/2$, the initial state evolves before the Ramsey measurement into
\begin{equation}
\ket{\psi(\tau)}
=
e^{-i\omega\tau\sigma_z/2}\ket{+}
=
\cos\frac{\omega\tau}{2}\ket{+}
-i\sin\frac{\omega\tau}{2}\ket{-}.
\end{equation}
Thus the pre-measurement weights in the $\sigma_x$ basis are
\begin{equation}
p_+(\tau)=\cos^2\frac{\omega\tau}{2},
\qquad
p_-(\tau)=\sin^2\frac{\omega\tau}{2}.
\end{equation}
Equivalently, the pre-measurement density matrix is
\begin{equation}
\rho_{\tau}^{-}
=
\ket{\psi(\tau)}\bra{\psi(\tau)}
=
\frac{1}{2}\left[
\mathbb{I}
+
\cos(\omega\tau)\sigma_x
+
\sin(\omega\tau)\sigma_y
\right].
\end{equation}

For a projective Ramsey measurement, $g\tau_A=\pi/2$, the nonselective measurement removes all coherences between $\ket{+}$ and $\ket{-}$. At $\omega\tau=\pi/2$, the two weights are equal, $p_+=p_-=1/2$, and the post-measurement state is therefore
\begin{equation}
\mathcal{M}_{\pi/(2g)}(\rho_{\tau}^{-})
=
\frac{1}{2}\ket{+}\bra{+}
+
\frac{1}{2}\ket{-}\bra{-}
=
\frac{\mathbb{I}}{2}.
\end{equation}
Thus one projective measurement maps the state directly to the fixed point. This gives the intuitive physical picture of the critical point in the projective-measurement limit: the qubit first evolves into an equal-weight superposition of the two measured eigenstates, and the following nonselective measurement converts this state into the maximally mixed state.

For a weak Ramsey measurement, $0<g\tau_A<\pi/2$, the measurement is not a projection. Instead, it only partially suppresses the components transverse to the measured $\sigma_x$ axis. Applying the measurement channel derived above to $\rho_\tau^-$ gives
\begin{equation}
\rho_{\tau}^{+}
=
\mathcal{M}_{\tau_A}(\rho_{\tau}^{-})
=
\frac{1}{2}\left[
\mathbb{I}
+
\cos(\omega\tau)\sigma_x
+
\cos(g\tau_A)\sin(\omega\tau)\sigma_y
\right].
\end{equation}
Therefore, the fastest approach to $\mathbb{I}/2$ is no longer determined by the equal-weight condition alone. The lower exceptional line selects the special pre-measurement state through
\begin{equation}
\sin(\omega\tau_{\mathrm{EP}})
=
\tan^2\frac{g\tau_A}{2},
\qquad
0<\omega\tau_{\mathrm{EP}}<\frac{\pi}{2}.
\end{equation}
On this lower exceptional line, the corresponding pre-measurement pure state is
\begin{equation}
\ket{\psi_{\mathrm{EP}}}
=
\sqrt{
\frac{1+\sqrt{1-\tan^4(g\tau_A/2)}}{2}
}\ket{+}
-i
\sqrt{
\frac{1-\sqrt{1-\tan^4(g\tau_A/2)}}{2}
}\ket{-}.
\end{equation}
The two components in this state become equal only in the projective limit $g\tau_A=\pi/2$. For weak measurements they are generally unequal, which explains why the weak-measurement fastest-mixing point has no simple equal-population projection picture. Instead, it is selected by the spectral condition for the SDM, where coherent rotation and partial measurement-induced dephasing cooperate most efficiently. We now verify the corresponding response sign explicitly.

\subsubsection{Survival probability}
We now specify the survival probability and the corresponding overlap coefficients for the closed-qubit model. We choose the initial state as $\rho_0=\ket{+}\bra{+}$, with $\ket{+}=(\ket{0}+\ket{1})/\sqrt{2}$, and denote the state after $m$ evolution--measurement cycles by $\rho_m=\Phi^m(\rho_0)$. The survival probability can be written in the vectorized operator space as
\begin{equation}
P(m)=\mathrm{Tr}[\rho_0\rho_m]
=\langle\!\langle \rho_0|\hat{\Phi}^{m}|\rho_0\rangle\!\rangle
=\sum_{i=1}^{4}d_i e^{m\lambda_i(\tau+\tau_A)}.
\end{equation}
Here $d_i$ is the overlap weight entering the survival probability, whereas $c_i$ in Sec.~S1A is the state-expansion coefficient $c_i=\langle\!\langle L_i|\rho_0\rangle\!\rangle$. The two coefficients are related by
\begin{equation}
d_i=\langle\!\langle \rho_0|R_i\rangle\!\rangle
\langle\!\langle L_i|\rho_0\rangle\!\rangle
=\langle\!\langle \rho_0|R_i\rangle\!\rangle c_i .
\end{equation}
For parameters away from the exceptional line, projecting $\rho_0$ onto the eigenmodes of the closed-qubit channel gives
\begin{equation}
d_1=\frac{1}{2},\qquad
d_{2,3}=\frac{1}{4}\left[
1\pm
\frac{\tan^2(g\tau_A/2)\cos(\omega\tau)}{\chi}
\right],\qquad
d_4=0,
\end{equation}
where
\begin{equation}
\chi=\sqrt{\tan^4(g\tau_A/2)-\sin^2(\omega\tau)}.
\end{equation}
Thus the $\Lambda_4$ branch does not contribute to the survival probability for this initial state. The long-time envelope is governed by the SDM selected from the $\Lambda_{2,3}$ branch. When this branch is a complex-conjugate pair, the imaginary part of $\lambda_{\mathrm{SDM}}$ produces damped oscillations, while the envelope decay rate is determined by $\Gamma=|\mathrm{Re}(\lambda_{\mathrm{SDM}})|$.


\subsection{Open qubit model}
We now extend the closed-qubit model by adding amplitude damping during the free-evolution stage. The coherent part of the dynamics is unchanged, while the free-evolution generator becomes $\mathcal{G}=\mathcal{L}_{B}+\mathcal{L}_{D}$, with $\mathcal{L}_{B}(\rho)=-i[H,\rho]$ and $H=\omega\sigma_z/2$. The dissipative part is taken as
\begin{equation}
\mathcal{L}_{D}(\rho)=\kappa\mathcal{D}[\sigma_-]\rho,
\end{equation}
where $\kappa$ denotes the amplitude-damping strength. The Ramsey measurement channel $\mathcal{M}_{\tau_A}$ is kept the same as in the closed-qubit model. Therefore the open-qubit evolution-measurement cycle is described by $\Phi_{\rm}(\tau,\tau_A)=\mathcal{M}_{\tau_A}e^{\mathcal{G}\tau}$.

The spectrum of this one-cycle channel can be obtained in the same way as in the closed-qubit model. The four channel eigenvalues are
\begin{equation}
\Lambda_1=1,\quad
\Lambda_{2,3}=
e^{-\kappa\tau/2}
\cos^2\frac{g\tau_A}{2}
\left[
\cos(\omega\tau)
\pm
\sqrt{
\tan^4\frac{g\tau_A}{2}
-
\sin^2(\omega\tau)
}
\right],
\quad
\Lambda_4=e^{-\kappa\tau}\cos(g\tau_A).
\end{equation}
Compared with the closed-qubit spectrum, amplitude damping only multiplies the conjugate-pair branch by $e^{-\kappa\tau/2}$ and the remaining Bloch branch by $e^{-\kappa\tau}$. The corresponding spectrum of the generator is obtained from $\lambda_i=\ln\Lambda_i/(\tau+\tau_A)$. If $\lambda_i^{(0)}$ denotes the spectrum of the generator in the closed-qubit model, then
\begin{equation}
\lambda_{2,3}=\lambda_{2,3}^{(0)}-\frac{\kappa\tau}{2(\tau+\tau_A)},
\qquad
\lambda_4=\lambda_4^{(0)}-\frac{\kappa\tau}{\tau+\tau_A}.
\end{equation}
Thus the damping shifts these decay modes along the negative real direction, with the $\Lambda_4$ branch shifted twice as strongly as the conjugate-pair branch. The EP condition is still determined by the vanishing of the square-root discriminant, namely $\tan^4(g\tau_A/2)=\sin^2(\omega\tau)$. Therefore the amplitude damping does not move the EP line, although it changes the relative decay rates of the modes. The SDM is the non-stationary mode with the largest channel-eigenvalue modulus, $|\Lambda_{\rm SDM}|=\max_{i=2,3,4}|\Lambda_i|$. In the weak-measurement regime $g\tau_A\ll1$, away from the trivial points $\sin(\omega\tau)=0$, the pair $\Lambda_{2,3}$ forms a complex-conjugate branch with
\begin{equation}
|\Lambda_{2,3}|=e^{-\kappa\tau/2}\sqrt{\cos(g\tau_A)},
\qquad
|\Lambda_4|=e^{-\kappa\tau}\cos(g\tau_A)=|\Lambda_{2,3}|^2.
\end{equation}
For any nontrivial decay, this gives $|\Lambda_{2,3}|>|\Lambda_4|$, so the SDM is the conjugate-pair branch $\Lambda_{2,3}$. In the additional weak-damping limit $\kappa\tau\ll1$, the corresponding decay rate is approximately $\Gamma_{\rm SDM}\simeq[\kappa\tau/2+(g\tau_A)^2/4]/(\tau+\tau_A)$. After crossing the EP line, the same branch splits into two real modes, and the SDM is selected by the real branch with the larger modulus.

\begin{figure}[H]
\centering
  \includegraphics[width=15cm]{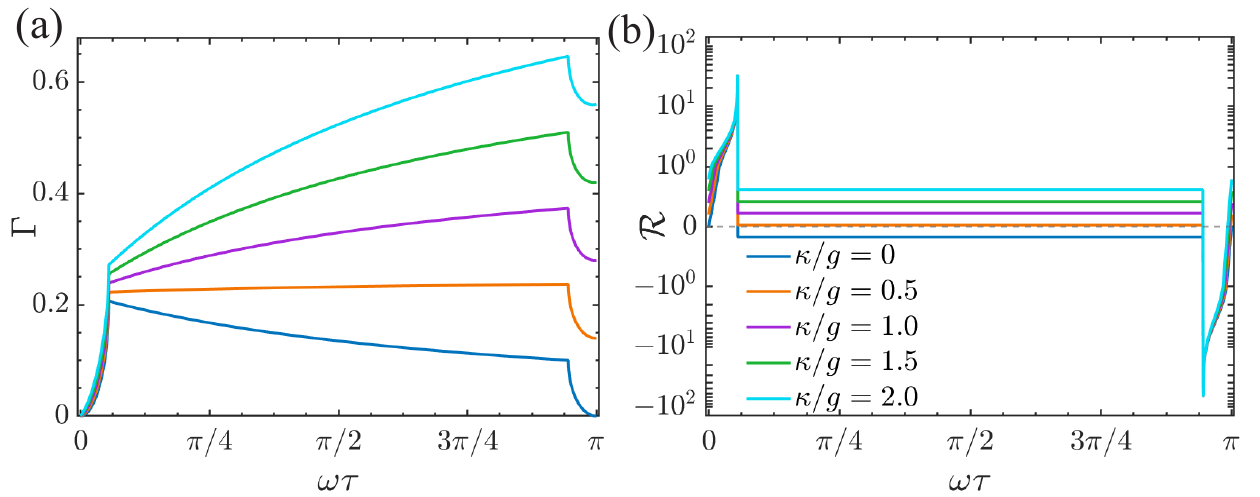}
  \caption{Spectral gap and response of the open-qubit discrete-time measurement model.
(a) Decay rate $\Gamma$ of the SDM as a function of $\omega\tau$ for different amplitude-damping strengths $\kappa/g$.
(b) The response $\mathcal R$ of the spectral gap for the same set of $\kappa/g$.
As $\kappa$ is varied, the sign of the plateau response changes, so the two EPs bounding the plateau alternately mark the boundaries between the Zeno and anti-Zeno responses.
(The fixed section is $g\tau_A=\pi/4$.)}
  \label{open_response}
\end{figure}

Using the SDM selected from the above spectrum, we now evaluate the Zeno and anti-Zeno response of the open-qubit model. Along a fixed-$g\tau_A$ cut, the EP condition can be satisfied at two values of $\omega\tau$. We call the interval between these two EPs the plateau region: in this interval the SDM is the complex-conjugate branch $\Lambda_{2,3}$, and the response obtained from this branch is independent of $\omega\tau$. For this SDM branch, the channel-eigenvalue modulus is $|\Lambda_{2,3}|=e^{-\kappa\tau/2}\sqrt{\cos(g\tau_A)}$, giving
\begin{equation}
\Gamma=
\frac{\kappa\tau/2-\ln\sqrt{\cos(g\tau_A)}}{\tau+\tau_A}.
\end{equation}
The resulting SDM decay rate along the fixed section $g\tau_A=\pi/4$ is shown in Fig.~\hyperref[open_response]{\ref*{open_response}(a)} for different damping strengths $\kappa/g$.
Therefore the response in the plateau region is
\begin{equation}
\mathcal{R}
=
\frac{\kappa\tau_A}{2}
+
\ln\sqrt{\cos(g\tau_A)}.
\end{equation}
The sign of this expression determines whether the plateau region exhibits a Zeno or anti-Zeno response. Setting $\mathcal{R}=0$ gives the critical damping strength
\begin{equation}
\frac{\kappa_c}{g}
=
-\frac{\ln[\cos(g\tau_A)]}{g\tau_A}.
\end{equation}
Thus, for $\kappa>\kappa_c$ the plateau region has $\mathcal{R}>0$ and displays a Zeno response, whereas for $\kappa<\kappa_c$ it has $\mathcal{R}<0$ and displays an anti-Zeno response (see Fig.~\hyperref[open_response]{\ref*{open_response}(b)}). This explains why the actual response boundary can move from one EP to the other as $\kappa$ is varied: the EP line itself is fixed by the square-root discriminant and is independent of $\kappa$, but the SDM hierarchy and the sign of $\mathcal{R}$ between the two EPs are changed by damping. In the closed-system case $\kappa=0$, the plateau response reduces to $\mathcal{R}=\ln\sqrt{\cos(g\tau_A)}<0$, so the corresponding plateau is always anti-Zeno.

\subsection{Extension to multi-qubit models}
We next consider an Ising model in a transverse field \cite{Quan2006} as a multi-spin environment to illustrate how the same spectral construction extends beyond the single-qubit examples. Instead of probing a single Bloch component, the Ramsey probe now couples to the collective transverse magnetization of the environment. For an $N$-spin cluster, we write
\begin{equation}
H_E^{(N)}=-J\left[\sum_{\langle j,k\rangle} Z_jZ_k+\lambda\sum_{j=1}^N X_j\right],
\qquad
H_{\rm int}^{(N)}=-J\delta Z_p\otimes B_N,\quad B_N=\sum_{j=1}^N X_j ,
\end{equation}
where $X_j$ and $Z_j$ are Pauli operators of the environmental spins, $Z_p$ is the probe operator, and $\langle j,k\rangle$ denotes nearest-neighbor bonds. Thus the two Ramsey branches shift the transverse field in opposite directions, in the same spirit as a transverse-field Ising environment subject to a weak conditional perturbation. With $\chi=J\delta\tau_A$ and Ramsey phase $A_\phi=0$, the nonselective probe readout induces $M_0=\cos(\chi B_N)$ and $M_1=\sin(\chi B_N)$ on the environment, so $\mathcal{M}_{\tau_A}(\rho)=M_0\rho M_0^\dagger+M_1\rho M_1^\dagger$. The one-cycle channel is therefore $\Phi(\tau,\tau_A)=\mathcal{M}_{\tau_A}\circ\mathcal{U}_\tau$, with $\mathcal{U}_\tau(\rho)=e^{-iH_E^{(N)}\tau}\rho e^{iH_E^{(N)}\tau}$. To obtain a closed analytic spectrum while retaining a genuine multi-spin environment, we specialize below to the minimal connected cluster with $N=2$, for which $H_E=-J[Z_1Z_2+\lambda(X_1+X_2)]$ and $B=B_2=X_1+X_2$. The environmental Hilbert space then has dimension $d=4$, and the channel has $16$ eigenvalues in Hilbert-Schmidt space.

For this two-spin cluster, the spectrum can be obtained by using the $X$ basis. The environmental Hilbert space splits into $\mathcal{H}_a=\operatorname{span}\{|++\rangle_x,|--\rangle_x\}$ and $\mathcal{H}_0=\operatorname{span}\{|+-\rangle_x,|-+\rangle_x\}$: the measurement operator $B$ has eigenvalues $2,-2$ in $\mathcal{H}_a$, but vanishes in $\mathcal{H}_0$. This separation has a simple meaning: $\mathcal{H}_a$ is visible to the collective Ramsey measurement, whereas $\mathcal{H}_0$ is a dark sector. In $\mathcal{H}_a$, the free Hamiltonian has energies $E_a^s=sJ\Omega$ with $\Omega=\sqrt{1+4\lambda^2}$ and $s=\pm$; in $\mathcal{H}_0$, the energies are $E_0^r=rJ$ with $r=\pm$. Defining $\beta_2=\cos(2\chi)$ and $\beta_4=\cos(4\chi)$, the complete channel spectrum is
\begin{equation}
\begin{aligned}
\operatorname{Spec}(\Phi)
&=
\operatorname{Spec}_{aa}\cup\operatorname{Spec}_{00}
\cup\operatorname{Spec}_{a0}\cup\operatorname{Spec}_{0a},\\
\operatorname{Spec}_{00}
&=
\left\{1,1,e^{-i2J\tau},e^{i2J\tau}\right\},\\
\operatorname{Spec}_{a0}
&=
\left\{\beta_2 e^{-i(E_a^s-E_0^r)\tau}\right\}_{s,r=\pm},
\qquad
\operatorname{Spec}_{0a}
=
\left\{\beta_2 e^{-i(E_0^r-E_a^s)\tau}\right\}_{s,r=\pm},\\
\operatorname{Spec}_{aa}
&=
\{1,\mu_1,\mu_2,\mu_3\}.
\end{aligned}
\end{equation}
Here the three nontrivial $aa$-sector eigenvalues are the roots of $\mu^3-T_a\mu^2+S_a\mu-\beta_4^2=0$, with
\begin{equation}
\begin{aligned}
T_a&=\beta_4\left[2C_a+n_x^2(1-C_a)\right]+C_a+n_z^2(1-C_a),\\
S_a&=\beta_4\left\{\beta_4\left[n_z^2+n_x^2C_a\right]+n_x^2(1-C_a)+2C_a\right\},
\end{aligned}
\end{equation}
where $C_a=\cos(2J\Omega\tau)$, $n_x^2=1/\Omega^2$, and $n_z^2=4\lambda^2/\Omega^2$. Physically, $\operatorname{Spec}_{a0}$ and $\operatorname{Spec}_{0a}$ describe coherences between a measurement-visible sector and the dark sector, while $\operatorname{Spec}_{aa}$ describes the competition between Ising-driven rotation and Ramsey-induced dephasing inside the visible sector. The branches relevant for the SDM comparison can be isolated as
\begin{equation}
\Lambda_c^{sr}=\beta_2 e^{-i(E_a^s-E_0^r)\tau},
\qquad
\Lambda_c^{rs,*}=\beta_2 e^{-i(E_0^r-E_a^s)\tau},
\qquad
\Lambda_{v,j}=\mu_j ,
\end{equation}
with $s,r=\pm$ and $j=1,2,3$. Thus all cross-sector coherence branches have the same modulus $|\Lambda_c^{sr}|=|\beta_2|$, whereas the three visible-sector branches have moduli $|\mu_j|$ determined by the cubic equation above.

\begin{figure}[htbp]
\centering
  \includegraphics[width=15cm]{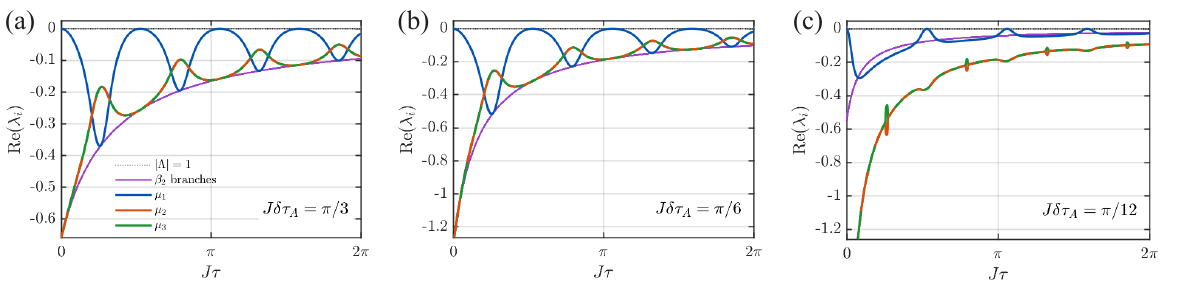}
  \caption{Effective generator spectra of the two-spin transverse-field Ising environment under collective Ramsey measurements.
Panels (a)--(c) show $\mathrm{Re}(\lambda_i)$ as a function of $J\tau$ for $J\delta\tau_A=\pi/3,\pi/6,\pi/12$, respectively. The $\beta_2$ branches denote the coherence modes between the $B=\pm2$ and $B=0$ subspaces with modulus $\beta_2$.}
  \label{muti_spectrum}
\end{figure}

The effective-generator spectrum follows from $\lambda_i^{\rm eff}=\operatorname{Log}\Lambda_i/(\tau+\tau_A)$. For the SDM-relevant branches, the decay rates are
\begin{equation}
\Gamma_c=-\frac{\ln|\beta_2|}{\tau+\tau_A},
\qquad
\Gamma_{v,j}=-\frac{\ln|\mu_j|}{\tau+\tau_A}\quad (j=1,2,3).
\end{equation}
The corresponding effective generator spectra are shown in Fig.~\hyperref[muti_spectrum]{\ref*{muti_spectrum}} for values of $J\delta\tau_A$. The multiple spectral crossings between decaying branches change the SDM hierarchy and mark the boundaries between the Zeno and anti-Zeno responses.
The phases of $\Lambda_c^{sr}$ are set by the Bohr frequencies $E_a^s-E_0^r$, so these modes describe oscillatory coherences damped by the collective Ramsey readout. The roots $\mu_j$ instead describe decay modes inside the measurement-visible sector itself, where the Ising rotation and the Ramsey backaction do not commute. The four modes in $\operatorname{Spec}_{00}$ have unit modulus because the Ramsey measurement cannot distinguish states in the $B=0$ dark sector; they are recurrent modes, not decay modes. Thus the SDM is selected only from the decaying branches,
\begin{equation}
|\Lambda_{\rm SDM}|
=
\max_{\Lambda\in\{\beta_2,\mu_1,\mu_2,\mu_3\},\,|\Lambda|<1}
|\Lambda|,
\qquad
\Gamma=-\frac{\ln|\Lambda_{\rm SDM}|}{\tau+\tau_A}.
\end{equation}
This form makes the physical mechanism transparent: the SDM switches according to whether the slowest relaxation is an inter-sector coherence suppressed by the collective Ramsey readout or an intra-sector decay mode produced by the combined action of free Ising dynamics and measurement backaction.

\section{Details for continuous-time measurement models}

\subsection{Superconducting-qubit–defect model}
We consider a superconducting qubit coupled to a microscopic two-level defect, with Pauli operators $\sigma_i$ and $\eta_i$ ($i=z,\pm$) for the two subsystems. Under the rotating-wave approximation, the coherent dynamics is described by $H_{qD}=(\omega_q/2)\sigma_z+(\omega_D/2)\eta_z+g_D(\sigma_+\eta_-+\sigma_-\eta_+)$, where $\omega_q$ and $\omega_D$ are the transition frequencies, $g_D$ is the exchange-coupling strength, and $\Delta=\omega_q-\omega_D$ is the detuning. Including intrinsic relaxation with rates $\gamma_q$ and $\gamma_D$, the free Liouvillian is
\begin{equation}
\mathcal{G}_{e}(\rho)
=
-i[H_{qD},\rho]
+
\gamma_q\mathcal{D}[\sigma_-]\rho
+
\gamma_D\mathcal{D}[\eta_-]\rho.
\end{equation}
The Hamiltonian term hybridizes the qubit and defect, while the dissipators relax them to their ground states.

Continuous measurement of the qubit is represented by the phase-damping generator $\mathcal{G}_m(\rho)=(\gamma_\phi/2)\mathcal{D}[\sigma_z]\rho$. It preserves qubit populations, suppresses qubit coherences, and provides the tunable measurement strength. The effective generator is therefore
\begin{equation}
\mathcal{G}_{\mathrm{eff}}(\rho)
=
-i[H_{qD},\rho]
+
\gamma_q\mathcal{D}[\sigma_-]\rho
+
\gamma_D\mathcal{D}[\eta_-]\rho
+
\frac{\gamma_\phi}{2}\mathcal{D}[\sigma_z]\rho.
\end{equation}
Its spectrum determines the SDM and the Liouvillian gap used for the continuous-time Zeno and anti-Zeno responses.

This generator picture is consistent with the Kofman--Kurizki spectral-overlap picture and the DISCo coarse-graining approach~\cite{Kofman2000,Kofman2001,Thorbeck2024}. There, measurement-induced dephasing broadens the qubit transition, and the defect-induced decay is obtained by averaging the loss spectrum $J(\omega)$ over the broadened qubit response,
\begin{equation}
\Gamma_{\mathrm{KK}}(\tilde{\omega}_q)
=
\gamma_q
+
\int d\omega\,
J(\omega)
F_{\gamma_\phi}(\omega-\tilde{\omega}_q),
\qquad
F_{\gamma_\phi}(\omega-\tilde{\omega}_q)
=
\frac{1}{\pi}
\frac{\gamma_\phi}
{\gamma_\phi^2+(\omega-\tilde{\omega}_q)^2}.
\end{equation}
Here $\tilde{\omega}_q$ is the readout-shifted qubit frequency; we focus on dephasing and set $\tilde{\omega}_q=\omega_q$. The DISCo treatment in Ref.~\cite{Thorbeck2024} derives this overlap formula from a coarse-grained master equation including both intrinsic dissipation and measurement-induced dephasing.

For a single dissipative defect, the loss spectrum entering the overlap formula is Lorentzian, $J_D(\omega)\propto g_D^2(\gamma_D/2)/[(\gamma_D/2)^2+(\omega-\omega_D)^2]$. The overlap between this defect spectrum and the broadened qubit response gives a contribution of the form $2g_D^2a/(a^2+\Delta^2)$. In the weak-coupling regime, the self-consistent evaluation around the decaying qubit-population branch yields $a=\gamma_\phi+\gamma_D/2-\gamma_q/2$. We next derive the same expression from the Liouvillian spectrum and identify it as the SDM spectral gap of $\mathcal{G}_{\mathrm{eff}}$.

\subsection{Perturbative derivation of the spectral gap}
Although the spectrum of $\mathcal{G}_{\mathrm{eff}}$ can be diagonalized directly, the full expressions are cumbersome. We therefore extract the weak-coupling spectral gap by treating $g_D$ perturbatively in the single-excitation sector spanned by $\ket{q}\equiv\ket{e}_q\ket{g}_D$ and $\ket{D}\equiv\ket{g}_q\ket{e}_D$. We use $X=(\rho_{qq},\rho_{DD},\rho_{Dq},\rho_{qD})^T$, with $\rho_{\alpha\beta}=\langle\alpha|\rho|\beta\rangle$; the first two components are populations and the last two are coherences.

In the basis $X$, the projected generator is represented by
\begin{equation}
\hat{\mathcal{G}}_{\mathrm{eff}}
=
\begin{pmatrix}
-\gamma_q & 0 & -ig_D & ig_D\\
0 & -\gamma_D & ig_D & -ig_D\\
-ig_D & ig_D & -A+i\Delta & 0\\
ig_D & -ig_D & 0 & -A-i\Delta
\end{pmatrix},
\end{equation}
where $A=\gamma_\phi+(\gamma_q+\gamma_D)/2$ and $\Delta=\omega_q-\omega_D$. The diagonal terms give the bare population and coherence decay rates, while the off-diagonal terms proportional to $g_D$ mix the two sectors. For $g_D=0$, the nonstationary eigenvalues are $\lambda_q^{(0)}=-\gamma_q$, $\lambda_D^{(0)}=-\gamma_D$, $\lambda_{Dq}^{(0)}=-A+i\Delta$, and $\lambda_{qD}^{(0)}=-A-i\Delta$. Under $\gamma_q<\gamma_D$ and $\gamma_q<A$, the SDM is the branch connected to $\lambda_q^{(0)}=-\gamma_q$. Its weak-coupling correction follows from
\begin{equation}
\begin{aligned}
\lambda\rho_{qq}
&=
-\gamma_q\rho_{qq}
+
ig_D(\rho_{qD}-\rho_{Dq}),\\
\lambda\rho_{DD}
&=
-\gamma_D\rho_{DD}
-
ig_D(\rho_{qD}-\rho_{Dq}),\\
\lambda\rho_{Dq}
&=
(-A+i\Delta)\rho_{Dq}
+
ig_D(\rho_{DD}-\rho_{qq}),\\
\lambda\rho_{qD}
&=
(-A-i\Delta)\rho_{qD}
+
ig_D(\rho_{qq}-\rho_{DD}).
\end{aligned}
\end{equation}
We normalize the eigenvector so that $\rho_{qq}$ is dominant. Then $\rho_{Dq}$ and $\rho_{qD}$ are induced at first order in $g_D$, whereas $\rho_{DD}$ enters only at second order. Eliminating the coherences around $\lambda_q^{(0)}=-\gamma_q$ gives $a=A-\gamma_q=\gamma_\phi+\gamma_D/2-\gamma_q/2$, with $(a-i\Delta)\rho_{Dq}\simeq -ig_D\rho_{qq}$ and $(a+i\Delta)\rho_{qD}\simeq ig_D\rho_{qq}$, or equivalently $\rho_{Dq}\simeq[-ig_D/(a-i\Delta)]\rho_{qq}$ and $\rho_{qD}\simeq[ig_D/(a+i\Delta)]\rho_{qq}$. Substitution gives $\rho_{qD}-\rho_{Dq}\simeq ig_D[1/(a+i\Delta)+1/(a-i\Delta)]\rho_{qq}=ig_D[2a/(a^2+\Delta^2)]\rho_{qq}$.
Therefore, the qubit-population equation gives the SDM eigenvalue
\begin{equation}
\lambda_{\mathrm{SDM}}
=
-\gamma_q
-
2g_D^2
\frac{a}{a^2+\Delta^2}
+
O(g_D^4).
\end{equation}
The correction is therefore a second-order population--coherence--population process, and the weak-coupling Liouvillian spectral gap is
\begin{equation}
\Gamma
=
-\mathrm{Re}(\lambda_{\mathrm{SDM}})
=
\gamma_q
+
2g_D^2
\frac{a}{a^2+\Delta^2}.
\label{eq:weak_coupling_gap}
\end{equation}
This is the weak-coupling spectral-gap formula used in the main text. It also matches the Kofman--Kurizki spectral-overlap result for a single dissipative defect~\cite{Thorbeck2024}, with $\gamma_D\equiv\gamma_{1,D}$ and $\Delta=\omega_q-\omega_D$. Thus $a/(a^2+\Delta^2)$ is the defect-overlap factor, while our formulation identifies the same readout-induced lifetime suppression or enhancement with the Liouvillian gap $\Gamma=-\mathrm{Re}(\lambda_{\mathrm{SDM}})$.


\subsection{Zeno-to-anti-Zeno response transition}
\begin{figure}[H]
\centering
  \includegraphics[width=15cm]{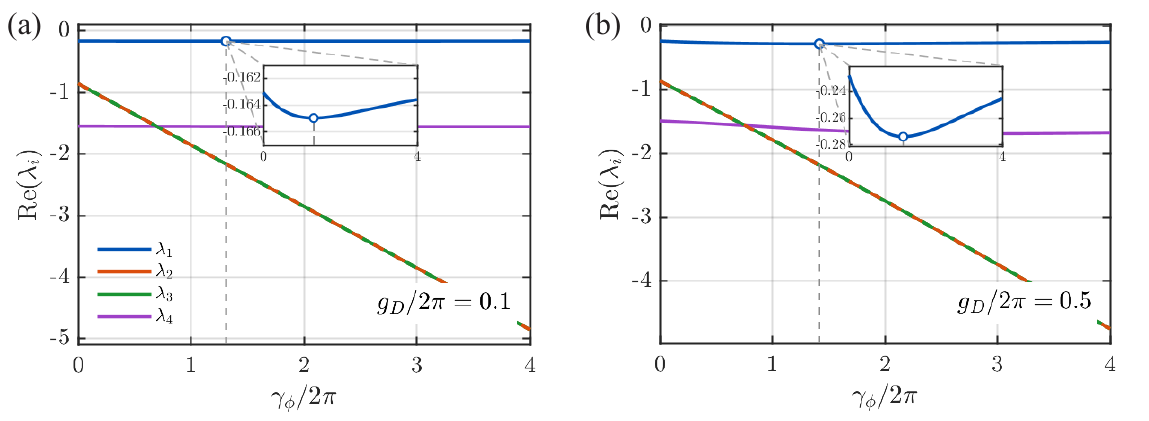}
  \caption{Effective-generator spectra of the superconducting-qubit--defect model.
Panels (a) (b) show $\mathrm{Re}(\lambda_i)$ as a function of $\gamma_\phi/2\pi$ for $g_D/2\pi=0.1,0.5~\mathrm{MHz}$, respectively.
The insets zoom in on the SDM branch $\lambda_1$; the open circles mark its extrema, where $\mathcal R_\phi=-\partial\Gamma/\partial\gamma_\phi$ changes sign.
($|\Delta|/2\pi=2~\mathrm{MHz}$, $\gamma_q/2\pi=0.16~\mathrm{MHz}$, and $\gamma_D/2\pi=1.55~\mathrm{MHz}$.)}
  \label{TLS_spectrum}
\end{figure}

From the weak-coupling spectral gap derived above, $\Gamma=\gamma_q+2g_D^2a/(a^2+\Delta^2)$ with $a=\gamma_\phi+\gamma_D/2-\gamma_q/2$, the continuous-time response $\mathcal{R}_\phi=-\partial\Gamma/\partial\gamma_\phi$ changes sign when the measurement-induced linewidth matches the detuning, namely $a=|\Delta|$. Thus the weak-coupling boundary between the Zeno and anti-Zeno responses is
\begin{equation}
\gamma_\phi^*
=
|\Delta|
-
\frac{\gamma_D}{2}
+
\frac{\gamma_q}{2}.
\end{equation}
This boundary reflects the competition between the effective linewidth of the qubit-defect transition and the detuning. However, it is obtained from the $O(g_D^2)$ correction to the eigenmode dominated by the qubit-excitation population, and therefore applies only in the weak qubit-defect coupling regime. As $g_D$ increases, the qubit and defect become strongly hybridized, the SDM is no longer a small perturbation of the uncoupled branch dominated by the qubit-excitation population, and the response boundary is shifted and distorted away from the weak-coupling line. This crossover is visible at the spectral level in Fig.~\hyperref[TLS_spectrum]{\ref*{TLS_spectrum}}. As $g_D$ increases, the extremum of the SDM branch shifts away from the weak-coupling prediction, indicating a displacement of the Zeno-to-anti-Zeno response boundary.

\begin{figure}[H]
\centering
  \includegraphics[width=15cm]{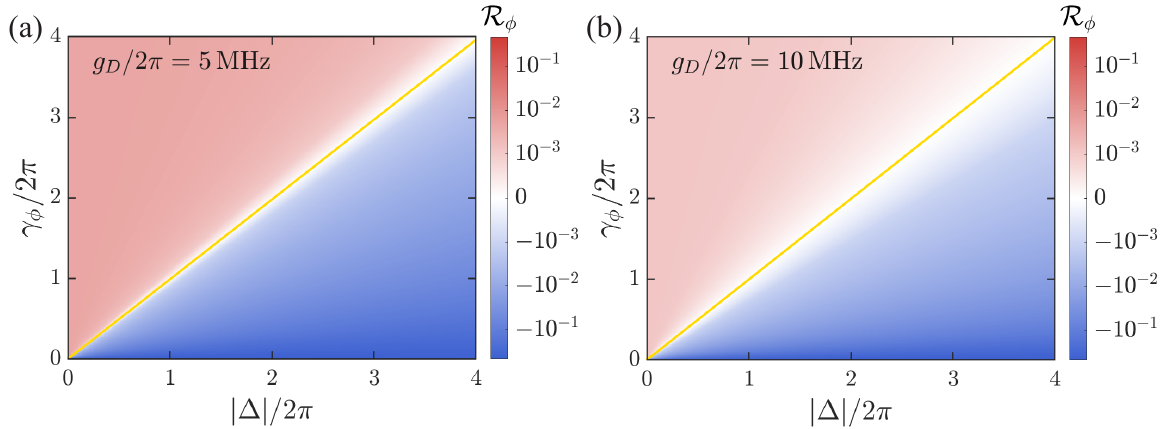}
  \caption{Response of the superconducting-qubit--defect model in the strong-coupling regime.
The heatmap shows the response $\mathcal R_\phi$ of the spectral gap in the $(|\Delta|/2\pi,\gamma_\phi/2\pi)$ plane.
Panels (a) and (b) correspond to stronger qubit-defect couplings $g_D/2\pi=5~\mathrm{MHz}$ and $10~\mathrm{MHz}$, respectively.
The yellow line denotes the strong-coupling boundaries between Zeno and anti-Zeno responses  $\gamma_\phi=|\Delta|$.}
  \label{TLS_response}
\end{figure}

The strong-coupling limit can be analyzed directly from the full projected generator $\hat{\mathcal{G}}_{\mathrm{eff}}$. We introduce the average relaxation rate $\bar{\gamma}:=(\gamma_q+\gamma_D)/2$ and the relaxation asymmetry $r:=(\gamma_D-\gamma_q)/2$,
so that $\gamma_q=\bar{\gamma}-r$, $\gamma_D=\bar{\gamma}+r$, and $A=\bar{\gamma}+\gamma_\phi$. To identify the SDM branch in the strong-coupling regime, we set $\lambda=-\bar{\gamma}+\epsilon$ and substitute it into the characteristic equation $\det(\lambda\mathbb{I}-\hat{\mathcal{G}}_{\mathrm{eff}})=0$. This gives
\begin{equation}
\begin{aligned}
0
=\,&
\epsilon^4
+
2\gamma_\phi\epsilon^3
+
\left(
4g_D^2+\gamma_\phi^2+\Delta^2-r^2
\right)\epsilon^2\\
&+
\left(
4\gamma_\phi g_D^2
-
2\gamma_\phi r^2
\right)\epsilon
-
r^2
\left(
\gamma_\phi^2+\Delta^2
\right).
\end{aligned}
\end{equation}
For $g_D\gg \gamma_\phi,|\Delta|,r$, the dominant $g_D^2$ terms are $4g_D^2\epsilon(\epsilon+\gamma_\phi)$. Hence two slow roots are located near $\epsilon\simeq0$ and $\epsilon\simeq-\gamma_\phi$, corresponding to $\lambda\simeq-\bar{\gamma}$ and $\lambda\simeq-\bar{\gamma}-\gamma_\phi$. The former has the larger real part for $\gamma_\phi>0$ and is therefore the SDM branch, while the remaining two roots are strongly oscillatory modes with imaginary parts of order $g_D$. Keeping the leading correction to the root near $\epsilon=0$, the characteristic equation reduces to $4\gamma_\phi g_D^2\epsilon-r^2(\gamma_\phi^2+\Delta^2)\simeq0$, which yields
\begin{equation}
\lambda_{\mathrm{SDM}}
\simeq
-\bar{\gamma}
+
\frac{r^2}{4g_D^2}
\frac{\gamma_\phi^2+\Delta^2}{\gamma_\phi}.
\end{equation}
Thus the strong-coupling spectral gap is
\begin{equation}
\Gamma
\simeq
\bar{\gamma}
-
\frac{r^2}{4g_D^2}
\frac{\gamma_\phi^2+\Delta^2}{\gamma_\phi}.
\end{equation}
The response boundary is determined by $\partial\Gamma/\partial\gamma_\phi=0$. Since $(\gamma_\phi^2+\Delta^2)/\gamma_\phi=\gamma_\phi+\Delta^2/\gamma_\phi$, this condition gives
\begin{equation}
\gamma_\phi^*
=
|\Delta|.
\end{equation}
Therefore, in the strong-coupling limit the boundary between the Zeno and anti-Zeno responses approaches a line passing through the origin with unit slope in the $(|\Delta|,\gamma_\phi)$ plane (see Fig.~\hyperref[TLS_response]{\ref*{TLS_response}(a)(b)}).

\bibliography{zeno2}